\documentclass[useAMS,usenatbib,usegraphicx]{mn2e}

\newcommand{\msolar}{M$_{\odot}$}

\usepackage{soul}

\PassOptionsToPackage{normalem}{ulem}
\usepackage{ulem}
\usepackage{xcolor}

\usepackage{subfigure}
\usepackage{multirow}
\usepackage{amssymb}

\title[The Type Ibn SN 2018cow CSM]{Signatures of Circumstellar Interaction in the Unusual Transient AT2018cow}
\author[O. D. Fox et al.]{Ori D. Fox$^{1,2}$, Nathan Smith$^{3}$\\
$^{1}$Space Telescope Science Institute, 3700 San Martin Drive, Baltimore, MD 21218, USA.\\
$^{2}$ofox@stsci.edu\\
$^{3}$Steward Observatory, 933 N. Cherry Ave., Tucson, AZ 85721, USA.
}
\begin{document}

\maketitle
\begin{abstract}

AT2018cow is a unique transient that stands out due to its relatively fast light-curve, high peak bolometric luminosity, and blue color.  These properties distinguish it from typical radioactively powered core-collapse supernovae (SNe).  Instead, the characteristics are more similar to a growing sample of Fast Blue Optical Transients (FBOTs).  Mostly discovered at hundreds of Mpc, FBOT follow-up is usually limited to several photometry points and low signal-to-noise spectra.  At only $\sim$60 Mpc, AT2018cow offers an opportunity for detailed followup. Studies of this object published to date invoke a number of interpretations for AT2018cow, but none of these specifically consider the interacting Type Ibn SN subclass.  We point out that while narrow lines do not dominate the spectrum of AT2018cow, as narrow Balmer lines typically do in SNe IIn, the narrow lines in AT2018cow may nevertheless be a mix of unresolved HII region emission and emission from slow, pre-shock CSM.  We compare AT2018cow to interacting SNe Ibn and IIn and find a number of noteworthy similarities, including light-curve rise and fall times, peak magnitude, X-ray light-curves, and spectroscopic properties.  In particular, the He I lines in AT2018cow closely resemble those in some examples of SNe Ibn or transitional SNe Ibn/IIn objects.  We therefore explore the hypothesis that CSM interaction in a relatively H-poor system might have some merit in explaining observed properties of AT2018cow, and we go on to consider progenitor implications for AT2018cow, FBOTs, and SNe~Ibn.
\end{abstract}


\section{Introduction}
\label{sec:intro}

The Astronomical Transient AT2018cow has been identified as a nearby analog to the rare class of Fast Blue Optical Transients (FBOTs; \citealt{perley18}).  The FBOT classification is defined by the light-curve's relatively fast rise and decline timescale (on the order of days), high peak bolometric luminosity ($>10^{44}$~erg), and blue color \citep[e.g.,][]{drout14,arcavi16,tanaka16,pursiainen18}.  The origin of the FBOT subclass remains ambiguous, but the rapid evolution is usually attributed to a small ejecta mass ($M_{\rm ej} < 1$~\msolar), while the energy source is considered to originate from shock interaction \citep[e.g.,][]{chevalier11,balberg11,ginzburg14,rest18} or a central compact object \citep[e.g.,][]{kasen10}.

The ATLAS survey discovered AT2018cow on June 16, 2018 in the dwarf star-forming galaxy CGCG 137-068.  At a presumed distance of $\sim$60 Mpc \citep{prentice18}, AT2018cow easily became the most nearby FBOT, allowing for detailed and extensive multi-wavelength followup \citep{prentice18,perley18,sandoval18,kuin18,margutti18}.  These analyses have proposed different interpretations of the progenitor, including a tidal disruption event (TDE) in an intermediate mass black hole (IMBH; \citealt{perley18}), a magnetar \citep{margutti18}, and the electron capture of a merged white dwarf \citep{lyutikov18}.  

\citet{perley18} considered a supernova (SN) origin for AT2018cow and performed a detailed comparison to many SN subclasses.  They found some similarities to the Type IIn subclass, but concluded that the Type IIn lines were more narrow than those observed in AT2018cow ($v\sim6000$ km/s versus a few hundred km/s in SN 1998S).  The broad lines in AT2018cow therefore led \citet{perley18} to suggest that the H and He emission lines originate in the fast moving ejecta and not the pre-shock circumstellar medium (CSM).  

\citet{margutti18}, however, measure line velocities of only $v\sim4000$~km/s, which may be problematic for the interpretation that these lines are emitted by freely expanding ejecta.  Not all SNe with a surrounding CSM have lines as narrow as SN 1998S.  For example, the Type Ibn SN 2006jc was dominated by intermediate lines of $1000 < v < 4000$~km/s \citep{foley07,pastorello07}.  The intermediate-width lines in SN 2006jc are attributed to accelerated CSM material swept up in the shock, but they may also originate in faster progenitor wind speeds from varying stellar types.  

The apparent lack of order 100 km/s lines would not, in itself, immediately rule out shock interaction or shock breakout in a wind. The narrowest lines from slow pre-shock gas can be absent from observations for a number of possible reasons.  (1) Spectra must have high enough resolution to separate the narrow and intermediate-width components. (2) The progenitor's mass loss must be slow.  Even moderately fast progenitor wind mass loss of a few 10$^3$ km s$^{-1}$ can still give rise to CSM interaction as long as it is slower than the forward shock speed. (3) Some non-spherical geometries can hide the narrow emission, depending on viewing angle. (4) The slow pre-shock CSM must be ionized and dense enough that its recombination emission (depending on $n_e^2$) can compete with the luminosity of the SN photosphere.  There is a wide parameter space where CSM interaction may still occur, but where these conditions may not be met to yield observable narrow lines.  Many events classified as Type IIn show only intermediate-width (and broad) lines, lacking the narrowest component (e.g., the Type IIn SN~2012ab; \citealt{bilinski18}, and references within).  See \citealt{smith17b} for a recent overview of interacting SNe (Types IIn and Ibn).

An important point, however, is that AT2018cow does show narrow lines.  While they are weaker than the narrow Blamer lines that dominate the spectra of typical SNe IIn, they may nevertheless be indicative of dense CSM.  These weak narrow lines in AT2018cow have been attributed to unresolved H II region emission, but we point out that some of the narrow lines, namely the He I lines, may indeed arise from slow pre-shock CSM, whereas the intermediate-width components may arise from post-shock gas.  In this paper, we therefore explore the hypothesis that CSM interaction may play an important role in the unusual observed properties of AT2018cow.  We provide a more detailed comparison between AT2018cow and other types of interacting SNe, including Type Ibn and the Type IIn SNe 1998S and PTF09uj.  Section \ref{sec:comparison} compares the light-curves, spectra, and rise-time to SNe Ibn and IIn.  Section \ref{sec:conclusion} provides a discussion and conclusion.

\section{Comparison to SN\lowercase{e} I\lowercase{bn} and II\lowercase{n}}
\label{sec:comparison}

The Type IIn subclass is characterized by narrow emission lines (mostly H and He) originating from a dense, pre-existing CSM \citep{schlegel90,smith17b}.  The Type Ibn subclass is similar to the SNe IIn but lacks significant H in the spectra \citep[][and those within]{hosseinzadeh17}.  The Type Ibn and IIn subclasses represent $<$1\% and $<$10\% of core-collapse SNe, respectively \citep{smith11,shivvers17}.

\subsection{Light-Curve}
\label{sec:lightcurve}

The light-curve of AT2018cow has several distinguishing properties \citep{prentice18,perley18}.  The rise increases $>$5 mag in $\sim$3.5 days and can be characterized by a $\sim$1.5 day rise to peak from half-max ($r$-band).  The $r$-band peak reaches a magnitude of -19.9.   The light-curve declines to half-max ($r$-band) in just 3 days and decreases to below an $r$-band mag of -16 by $\sim$25 days post-discovery.  This behavior is quite extreme when compared to most non-interacting SN subclasses \citep{perley18}. Those authors did not undertake a comparison to Type Ibn events, which tend to show quicker declines than SNe IIn, and there is also a wide diversity among SNe IIn that was not considered.

\begin{figure}
\centering
\includegraphics[width=3.5in]{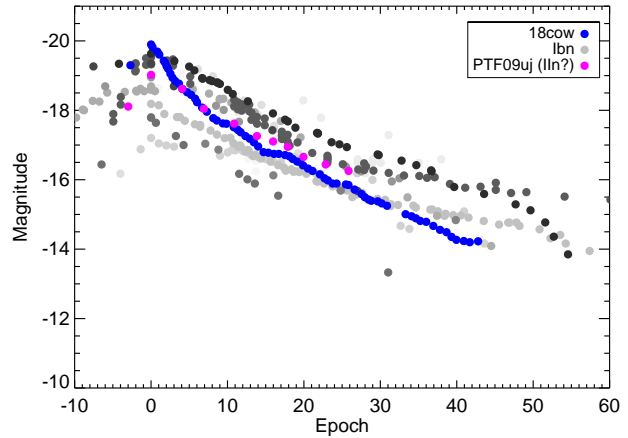}
\caption{Light-curve of AT2018cow (blue) compared to other SNe Ibn (shades of grey) and the Type IIn PTF09uj (magenta).}
\label{fig1} 
\end{figure}

Figure \ref{fig1} compares the light-curve of AT2018cow published by \citet{perley18} to SNe Ibn compiled from \citet{hosseinzadeh17}, including PTF11rfh, iPTF15ul, iPTF15akq, SN 1999cq, SN 2005la, SN 2006jc, SN 2010al, PS1-12sk, LSQ12btw, LSQ13ccw, SN 2014av, SN 2014bk, and ASASSN-15ed.  \citet{hosseinzadeh17} show that unlike SNe IIn, whose light-curves can vary \citep{kiewe12}, SNe Ibn tend to have fast, bright ($r\sim$-18 to -20 at peak), and relatively homogenous, well-defined light curves.  Most SNe IIn evolve on longer timescales, although we do include a comparison with the unique SNe IIn(?\footnote{PTF09uj was originally classified as a SNe IIn, but the progenitor and subsequent classification remains debated \citep{ofek10}.}) PTF09uj \citep{ofek10}.  AT2018cow qualitatively has a similar light-curve to these SNe, with its peak is only slightly brighter than most of the other targets.  Note however, that several SNe Ibn including the prototypical case of SN2006jc, were not discovered early enough to characterize the peak.  AT2018cow declines more rapidly than the other transients over the first few days after peak, but after that does not have any distinguishing light-curve features.

\begin{figure*}
\centering
\includegraphics[width=3.25in]{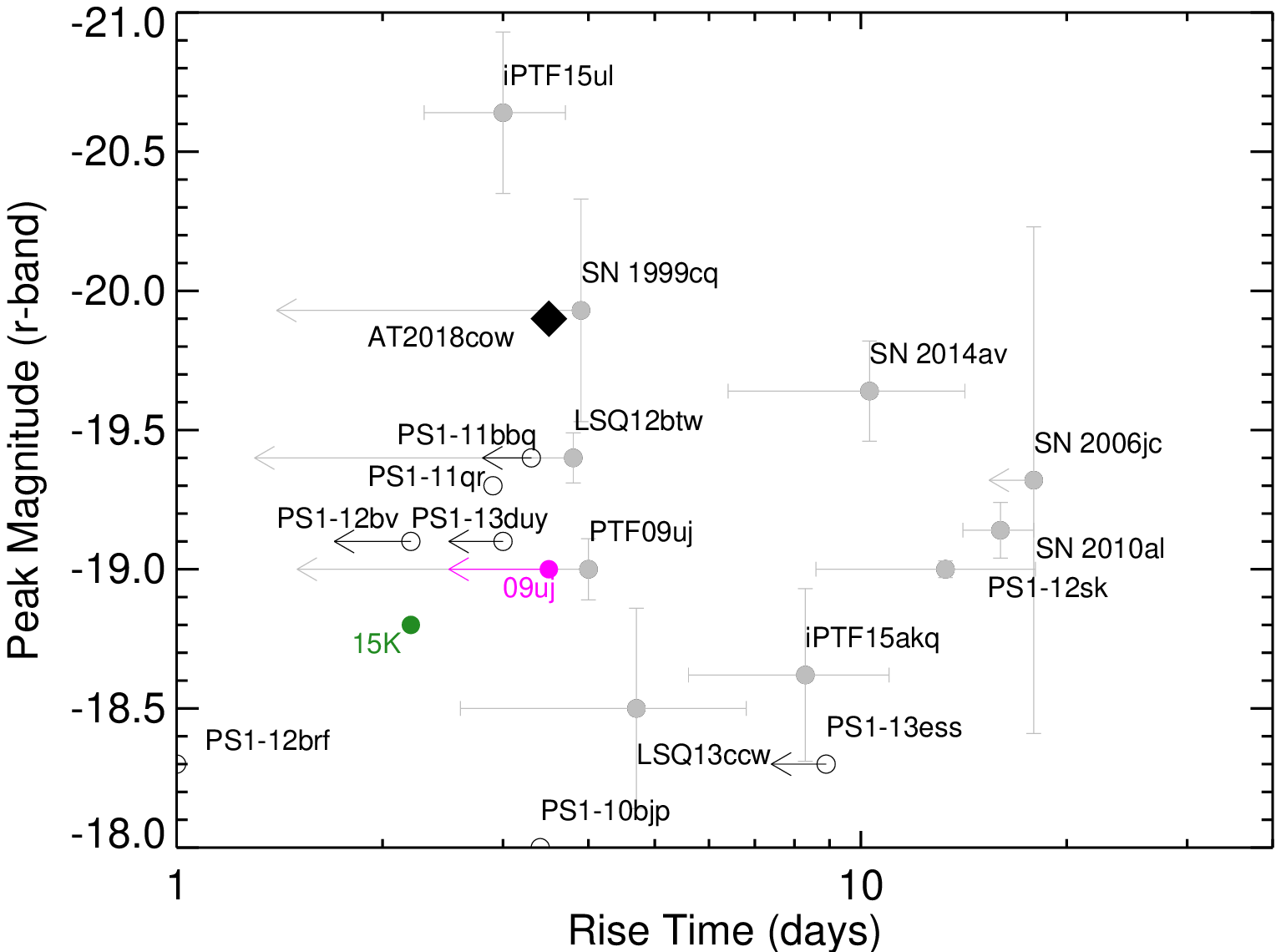}
\includegraphics[width=3.25in]{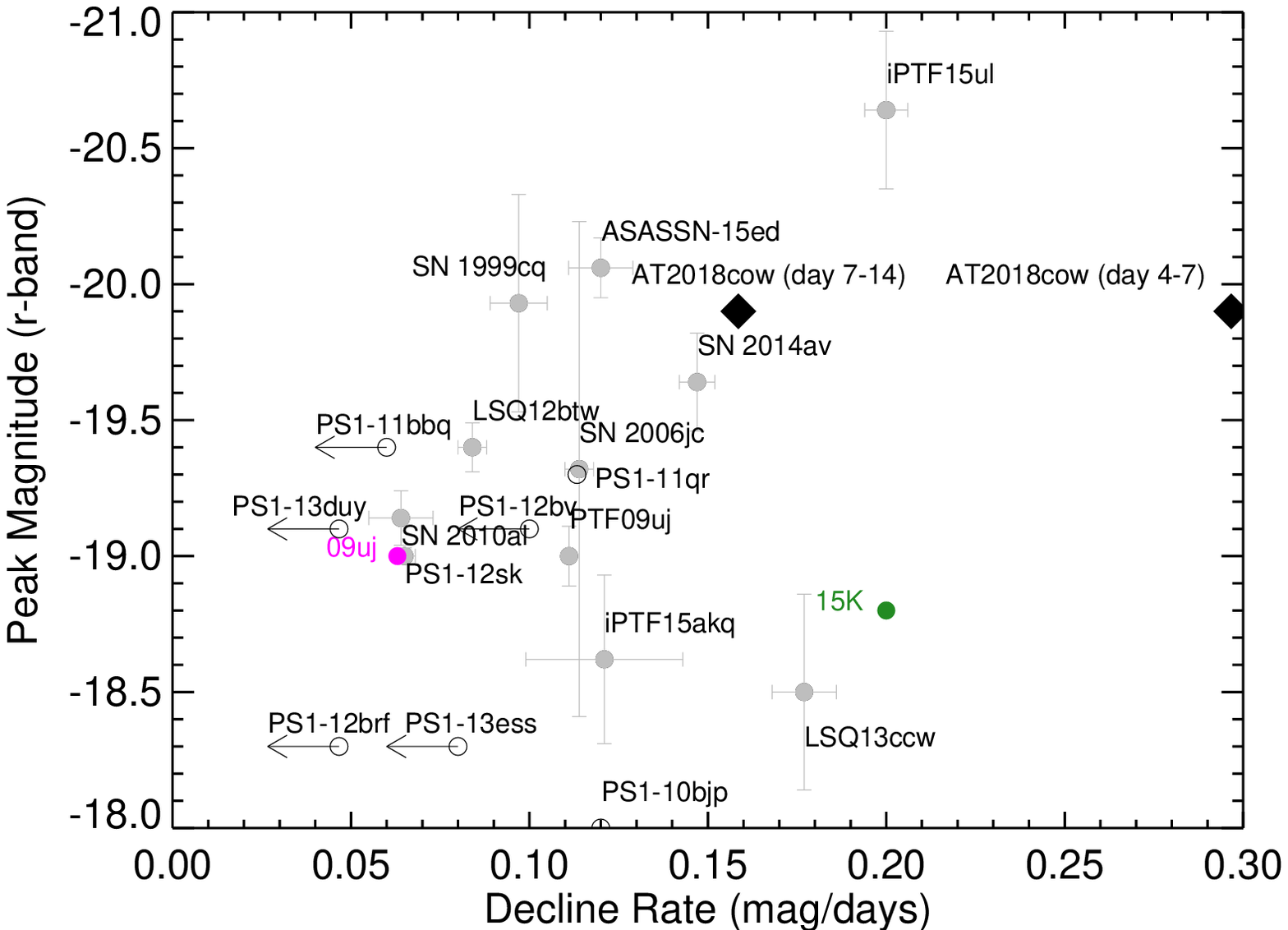}
\caption{Phase space plots.  (left) Peak luminosity versus rise time for AT2018cow (black) compared to SNe Ibn (grey), PTF09uj (magenta), Kepler SN 2015K (green), and the Pan-STARRS FBOT sample (open black circles). (right) Peak luminosity versus decline rate for AT2018cow (black) compared to SNe Ibn (grey), PTF09uj (magenta), Kepler SN 2015K (green), and the Pan-STARRS FBOT sample (open black circles).}
\label{fig2} 
\end{figure*}

\subsection{Phase Space}
\label{sec:phasespace}

\citet{margutti18} show in their figure 1 a plot of transient peak luminosity versus rise time.  In addition to AT2018cow, the plot also includes super-luminous SNe (SLSNe), SNe Ibc, SNe IIP/L/n, and previously observed FBOTs.  AT2018cow stands apart from these populations objects due to its high luminosity and short rise time. 

Elaborating on the \citet{margutti18} comparison, Figure \ref{fig2} plots the peak luminosity versus both rise time and decline rate for the SNe Ibn sample in Figure \ref{fig1}, where the values were obtained from table 4 in \citet{hosseinzadeh17} and are all corrected for extinction.  For comparison, we also include the Pan-STARRS FBOT sample, where the values were obtained from table 4 in \citet{drout14}, and the Kepler SN 2015K \citep{rest18}.  Interacting SNe can have long rise times if they have denser and more opaque inner winds \citep{smithmccray07,chevalier11}, so we again do not include those in this analysis except for PTF09uj.  Compared to other SNe Ibn (and even the Pan-STARRs FBOTs), the phase space occupied by AT2018cow is not a clear outlier.  In fact, SN 1999cq and iPTF15ul are even more luminous and have even faster rise times.  Also, as pointed out in Figure \ref{fig1}, AT2018cow declines slightly more rapid than the other transients from peak over the first few days, but after that the decline rate is more typical of other SNe Ibn.  This behavior is not reminiscent of some SNe with brief CSM interaction \citep[e.g.][]{morozova18}.  Unlike SNe IIP or IIb, the normal underlying light curve in this case is the decline of a stripped envelope SNe.

\subsection{Narrow and Intermediate Lines}
\label{sec:narrow}

\begin{figure*}
\centering
\includegraphics[width=2.15in]{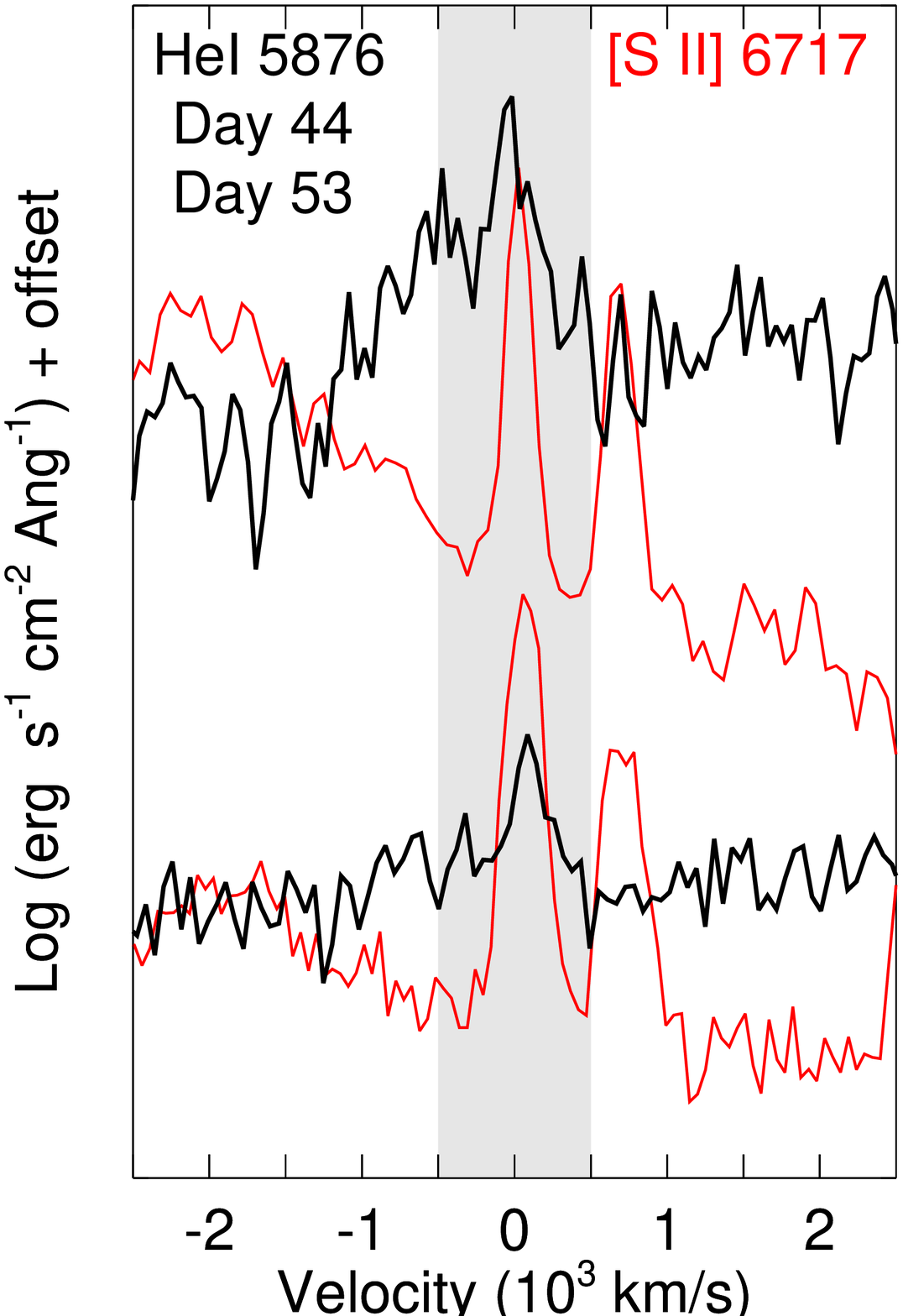}
\includegraphics[width=2.15in]{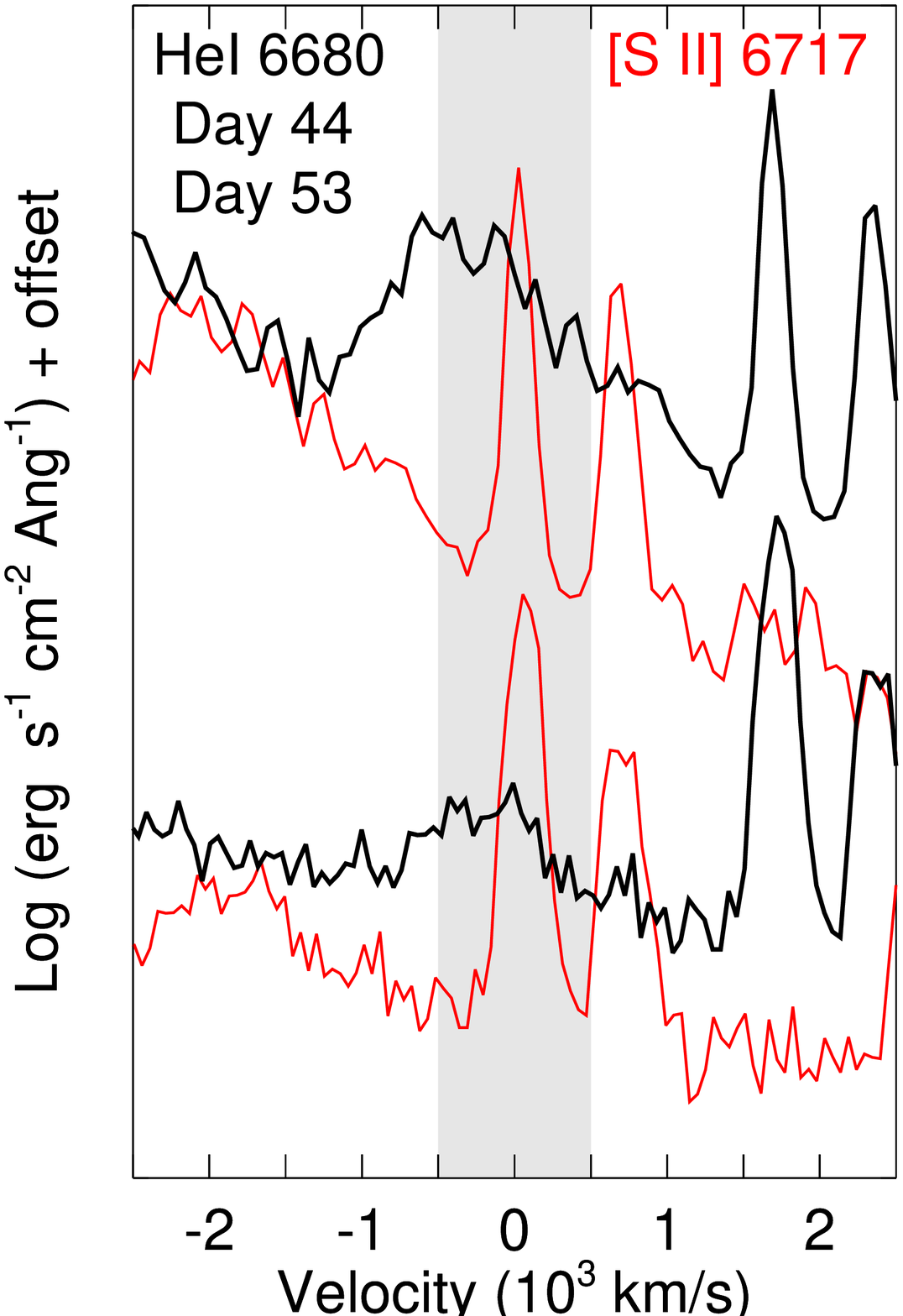}
\includegraphics[width=2.15in]{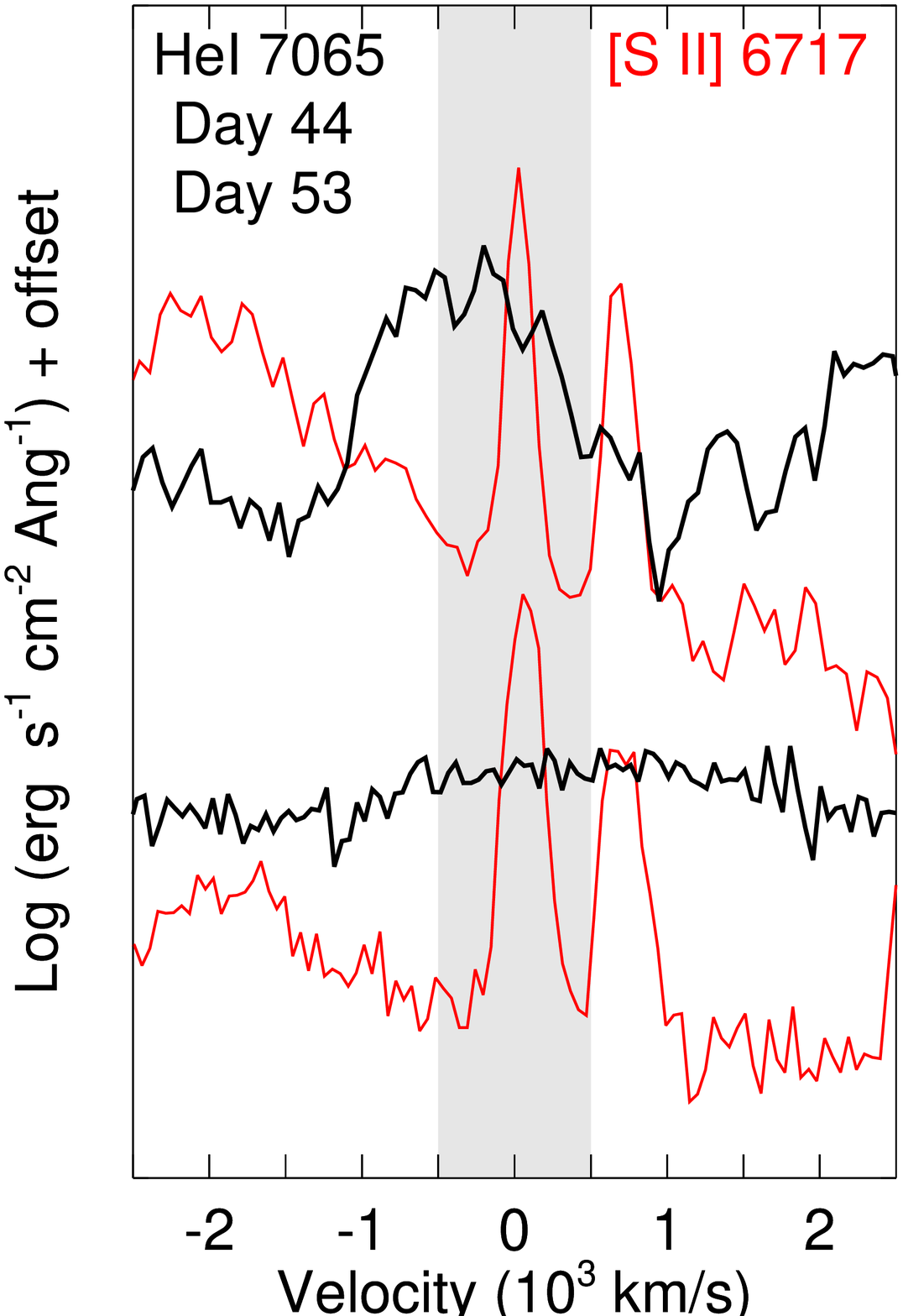}\\
\includegraphics[width=2.15in]{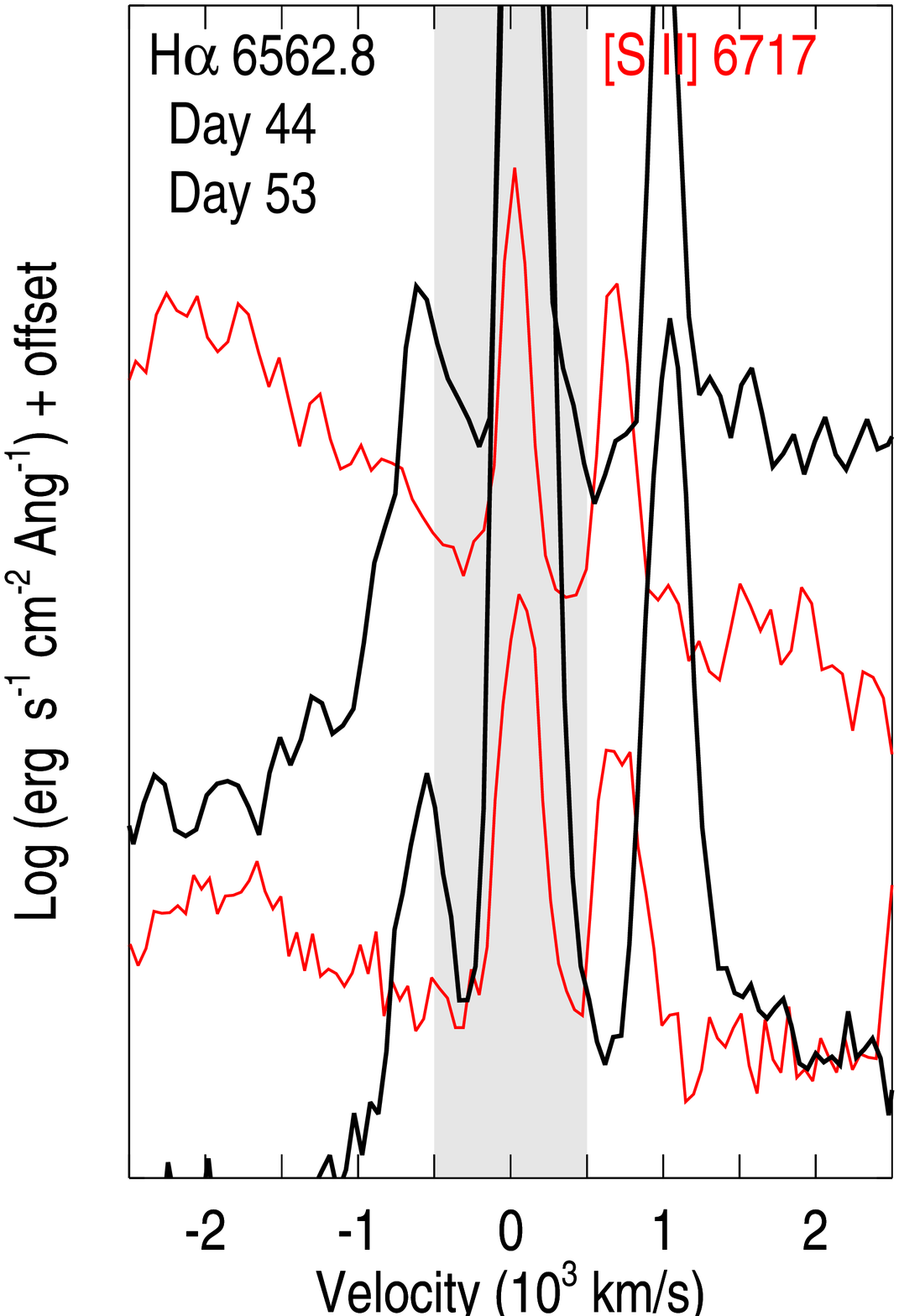}
\includegraphics[width=2.15in]{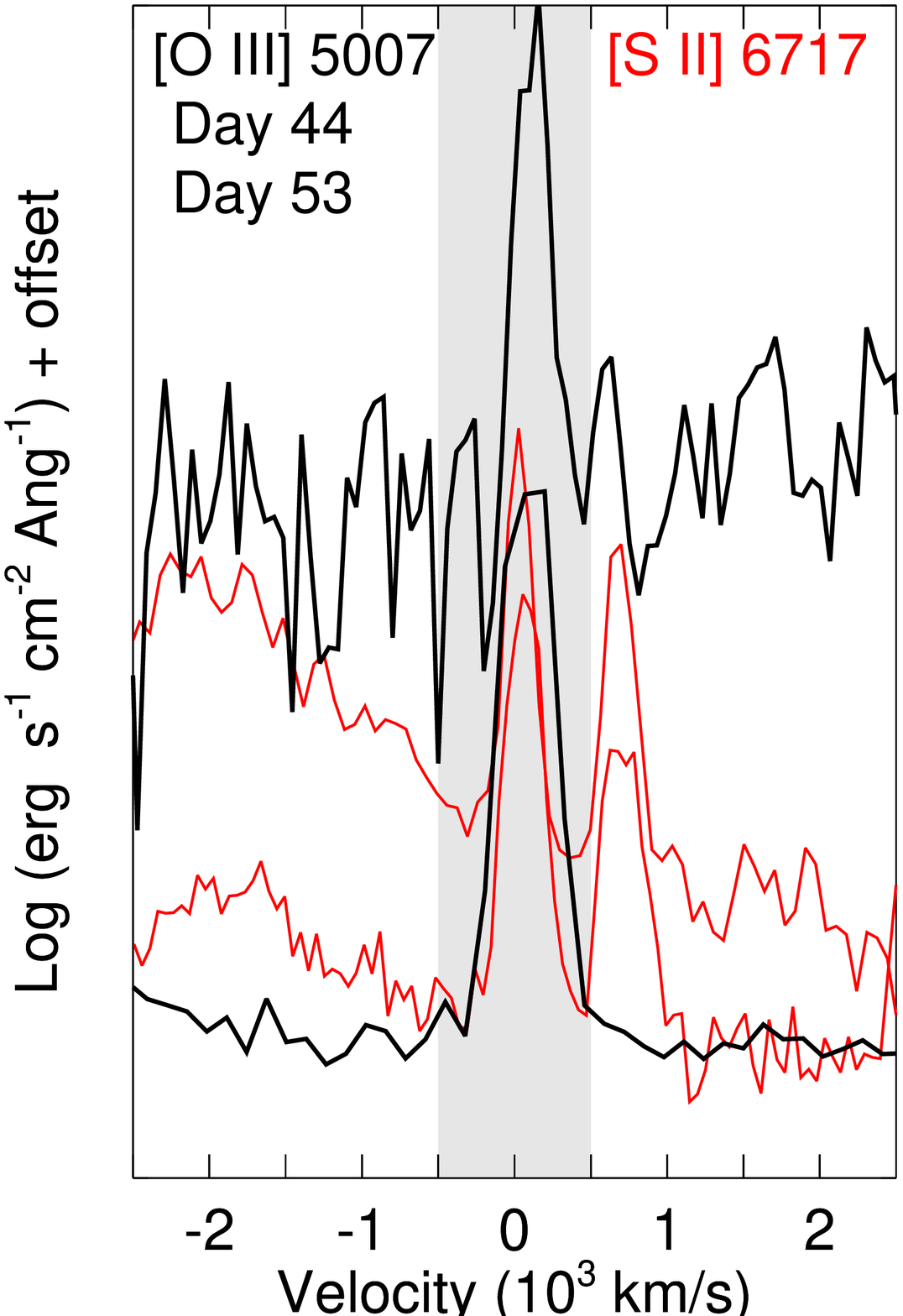}
\caption{Velocity plots of the helium, H$\alpha$, and [O III].  Overplotted (dotted, red lines) for comparison is the [S II] 6717 line originating from the underlying galaxy.  The shaded grey region highlights $\pm$500 km/s.}
\label{fig_he1vel} 
\end{figure*}



Figure \ref{fig_he1vel} plots the Helium, H$\alpha$, and [O III]~lines in velocity space for the day 44 and 53 spectra of AT2018cow from \citet{perley18}.  Overplotted for comparison is the [S II] 6717 line, which is quite narrow and we know originates from the underlying galaxy.  The day 44 spectrum was obtained with Gemini/GMOS using a 0.5\arcsec~slit and B600+R400 gratings, which have a resolving power of 1688 at 4610 \AA \, and 1918 at 7640 \AA, respectively.  This yields a resolution of $\sim$2.7 \AA~(or $\sim$175 km/s).  The day 53 spectrum was obtained with Keck/LRIS using a 1.0\arcsec~slit and the B400/3400 and R400/8500 grisms, which have a resolving power of 400 and a resolution of $\sim$6.9 \AA~(or $\sim$400 km/s).

As noted by \citet{perley18} and \citet{margutti18}, both narrow and intermediate components are apparent in the spectra.  While the most narrow lines have been attributed to the underlying galaxy, we note that the He I lines in the day 44 spectrum of AT2018cow are relatively strong compared to SN 1998S.  These lines seem too strong to be associated with an underlying HII region, judging by their strength compared to the narrow [O~{\sc iii}] lines, for example.  We measure a He I $\lambda$5876 to [O~{\sc iii}] $\lambda$5007 ratio of 0.84$\pm$0.03 and 0.87$\pm$0.04 on days 44 and 53, respectively.  The ratio observed in H~{\sc ii} regions, even with early O-type stars, is commonly $<$0.1 \citep[e.g.,][]{peimbert00,smith04,esteban04}.  Furthermore, the He I narrow lines are noticeably broader than the [O~{\sc iii}] 5007 and [S~{\sc ii}] 6723 lines that do likely originate in an H~{\sc ii} region.  The top, middle panel of Figure \ref{fig_he1vel} (He I 6680) illustrates this best.  The [S~{\sc ii}] $\lambda\lambda$6717,6731 lines (just to the right of the He I 6680 line) are clearly narrower than the  He I $\lambda$6680 line itself in both spectra.  We also fit a gaussian to these lines and find the He I $\lambda$6680 lines have a FWHM of 732$\pm$30 and 734$\pm$30 km/s on days 44 and 53, respectively.  By contrast, we find the [O~{\sc iii}] 5007 have a FWHM of 280$\pm$20 and 306$\pm$20 km/s on days 44 and 53, respectively.

\citet{perley18} and \citet{margutti18} also point out intermediate (on order of several thousand km/s) lines.  Compared to \citet{perley18}, \citet{margutti18} show in their figure 2 that the intermediate H and He lines have associated velocities of $\sim$4000 km/s, which is approaching speeds that are typically a bit too slow for ejecta of such an energetic explosion at such early times.  These velocities are often observed in the post-shocked CSM of many SNe IIn \citep{kiewe12,smith17}.  We also note that stars can have a range of pre-explosion wind speeds, ranging from red super giants (RSGs) with tens of km/s \citep[e.g.,][]{smith09ip} to luminous blue variables (LBVs) with few hundred km/s, 
to stripped Wolf–Rayet (W-R) stars with thousands of km/s (see \citealt{smith14} for a review of massive star mass loss properties). In other words, stars have a wide range of line widths and are not constrained to the narrow line widths of SN 1998S.  Furthermore, as noted in the Introduction, narrow lines from a slow pre-shocked CSM may not be observed for a variety of reasons.

\subsection{Spectroscopic Comparisons}
\label{sec:spectra}

\begin{figure*}[b]
\centering
\includegraphics[width=7.5in]{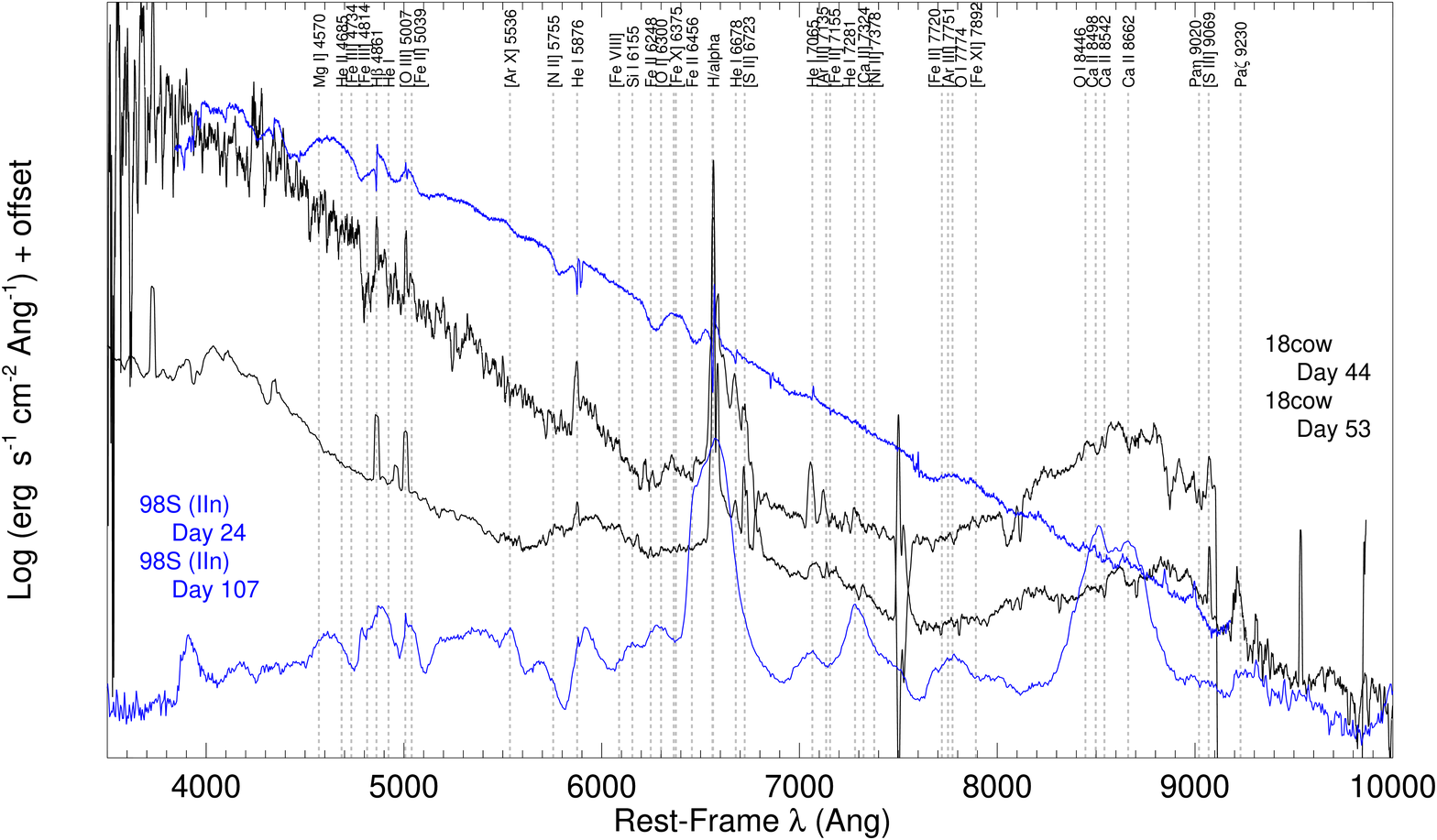}
\caption{Comparison of AT2018cow (black) to the Type IIn SN 1998S (blue).  Many high-ionization spectral lines typically seen in interacting SNe are labeled, but not all are necessarily identified in each spectrum.}
\label{fig3} 
\end{figure*}

\begin{figure*}
\centering
\includegraphics[width=7.5in]{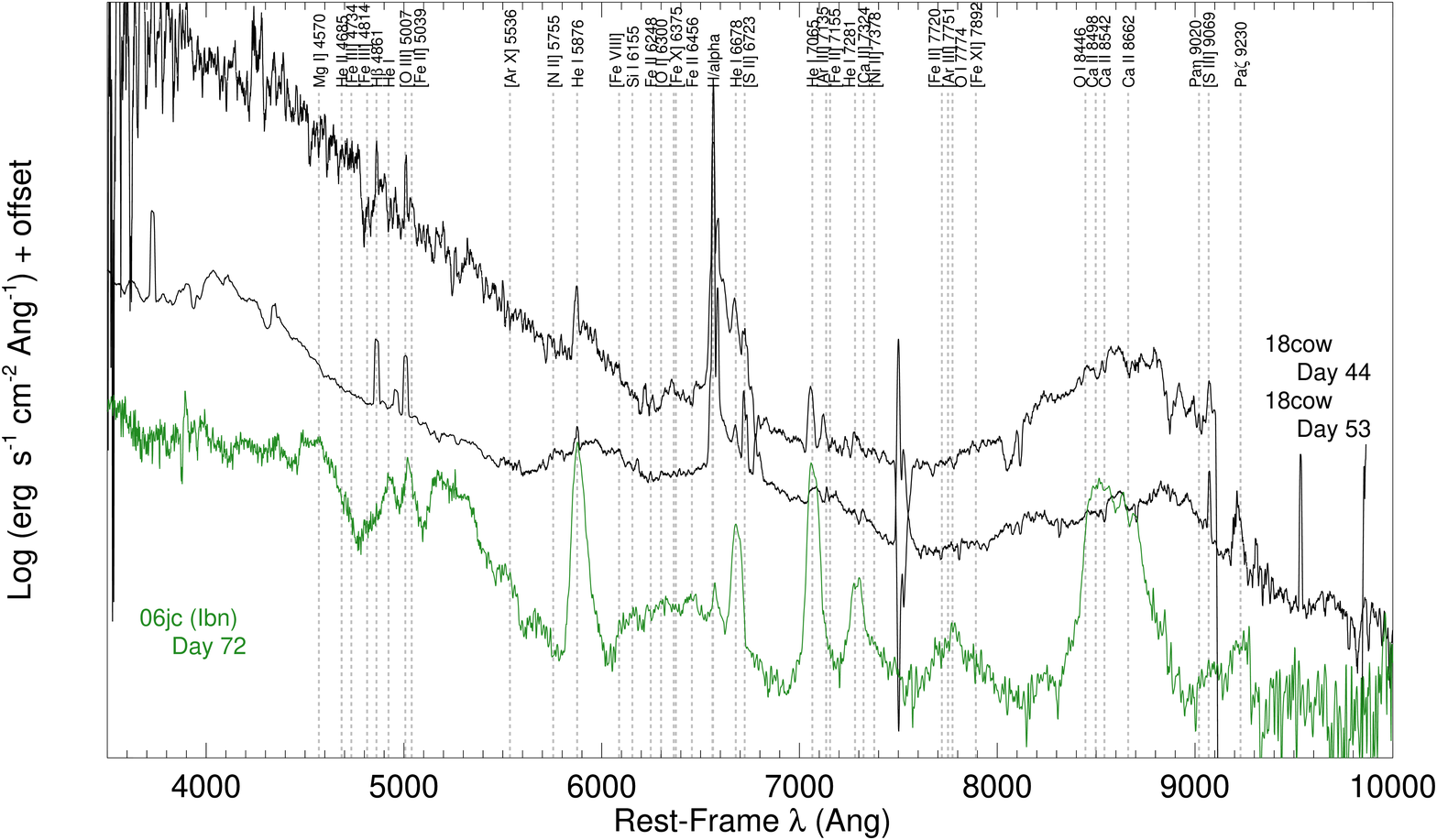}
\includegraphics[width=7.5in]{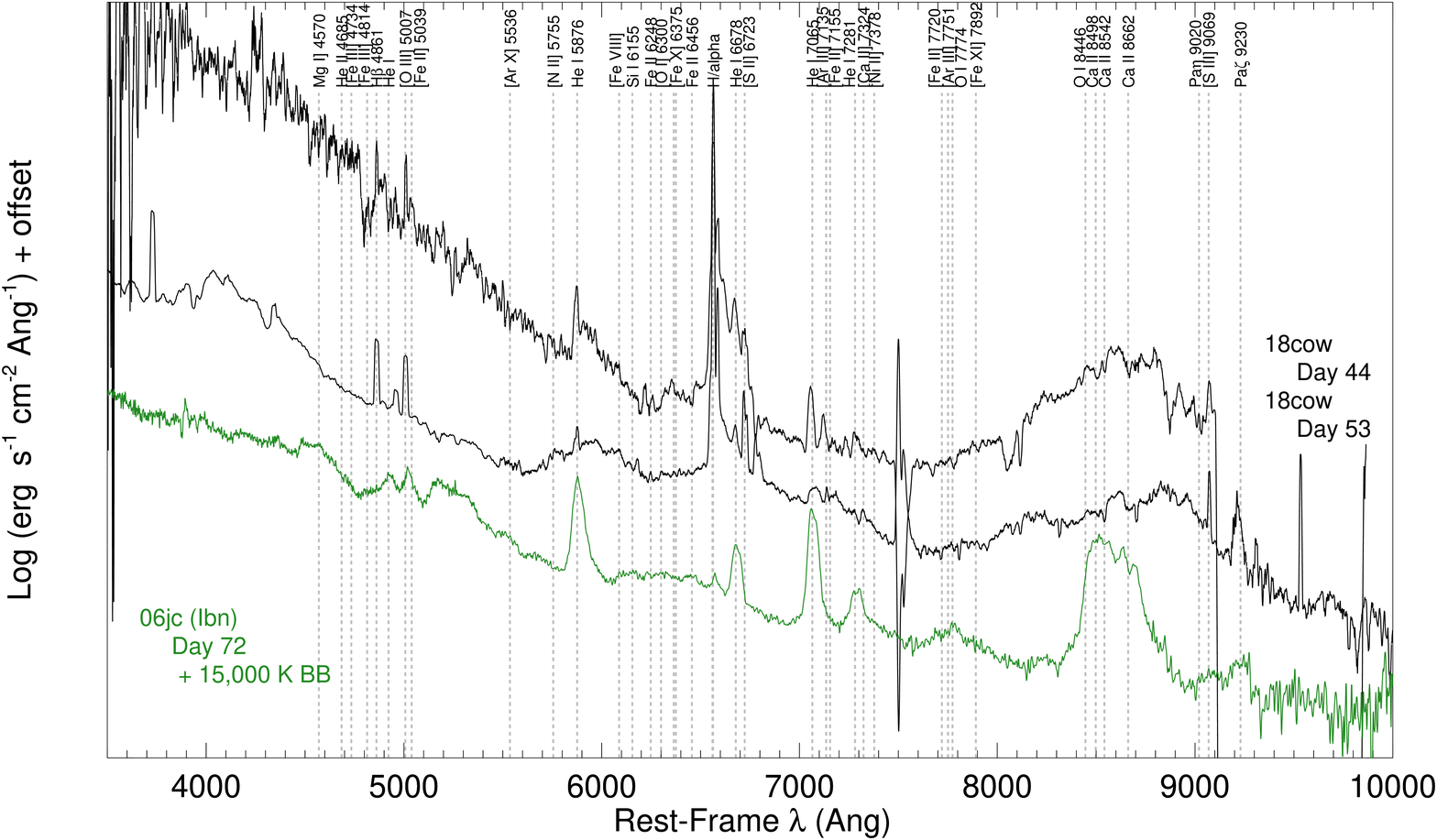}
\caption{{\bf (Top)} Comparison of AT2018cow (black) with the Type Ibn SN 2006jc (green).  {\bf (Bottom)} The spectrum of SN 2006jc is combined with a 15,000 K blackbody spectrum that represents the optically thick photosphere.  Many high-ionization spectral lines typically seen in interacting SNe are labeled, but not all are necessarily identified in each spectrum.}
\label{fig4} 
\end{figure*}

\begin{figure*}
\centering
\includegraphics[width=7.5in]{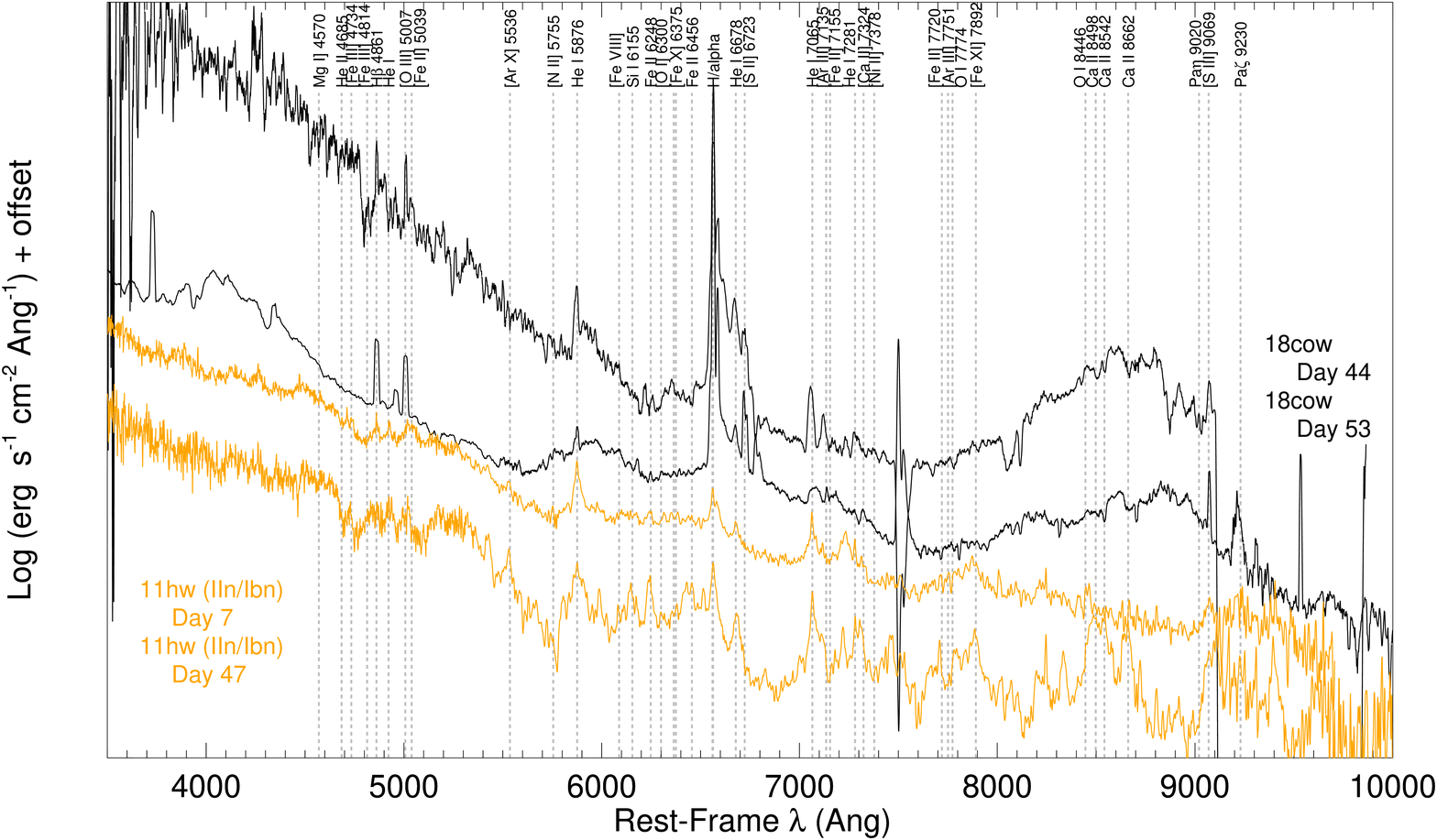}
\caption{Comparison of AT2018cow (black) with the Type IIn/Inb SN 2011hw (orange).  Many high-ionization spectral lines typically seen in interacting SNe are labeled, but not all are necessarily identified in each spectrum.}
\label{fig5} 
\end{figure*}

\begin{figure*}
\centering
\includegraphics[width=7.5in]{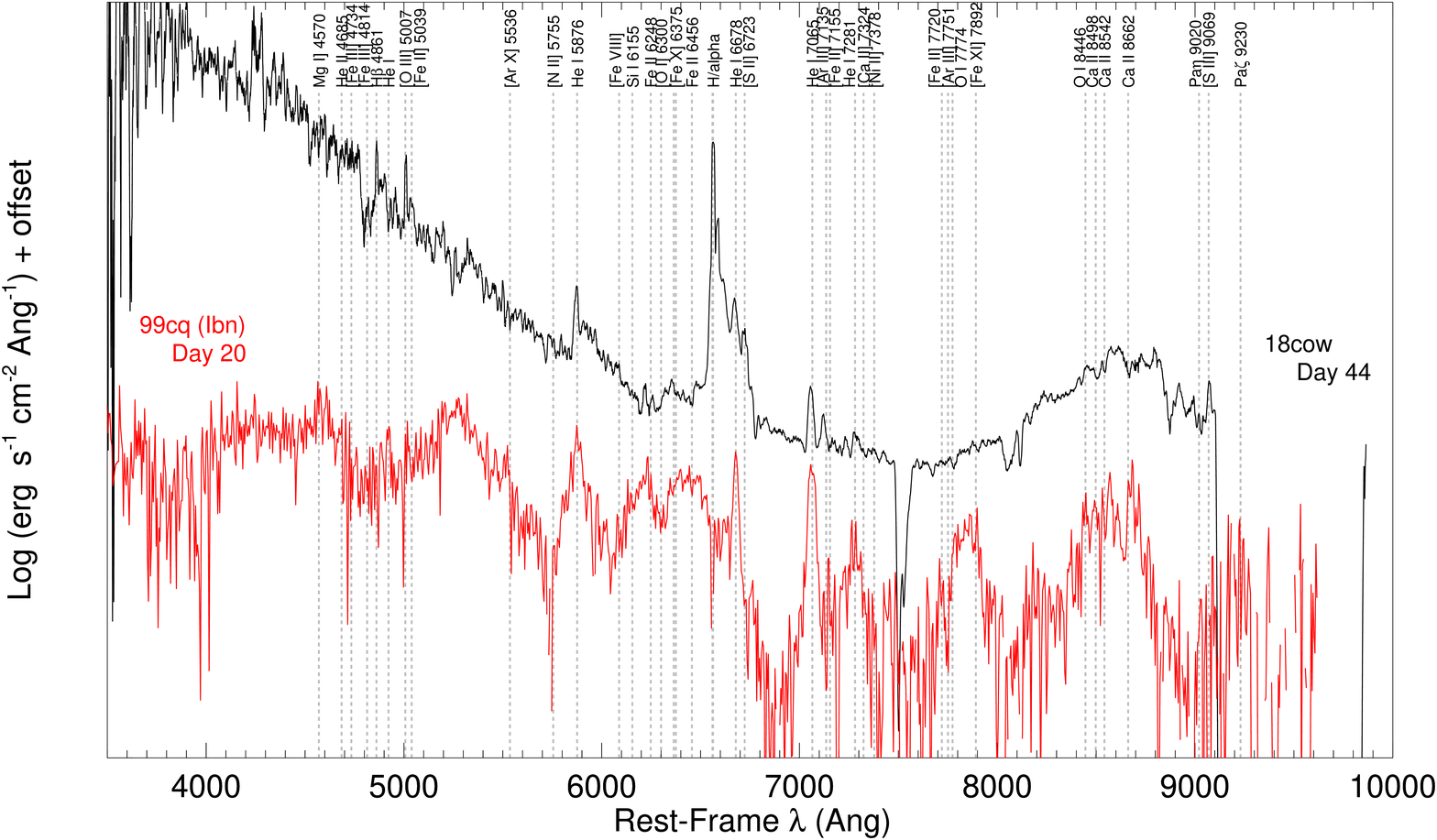}
\includegraphics[width=7.5in]{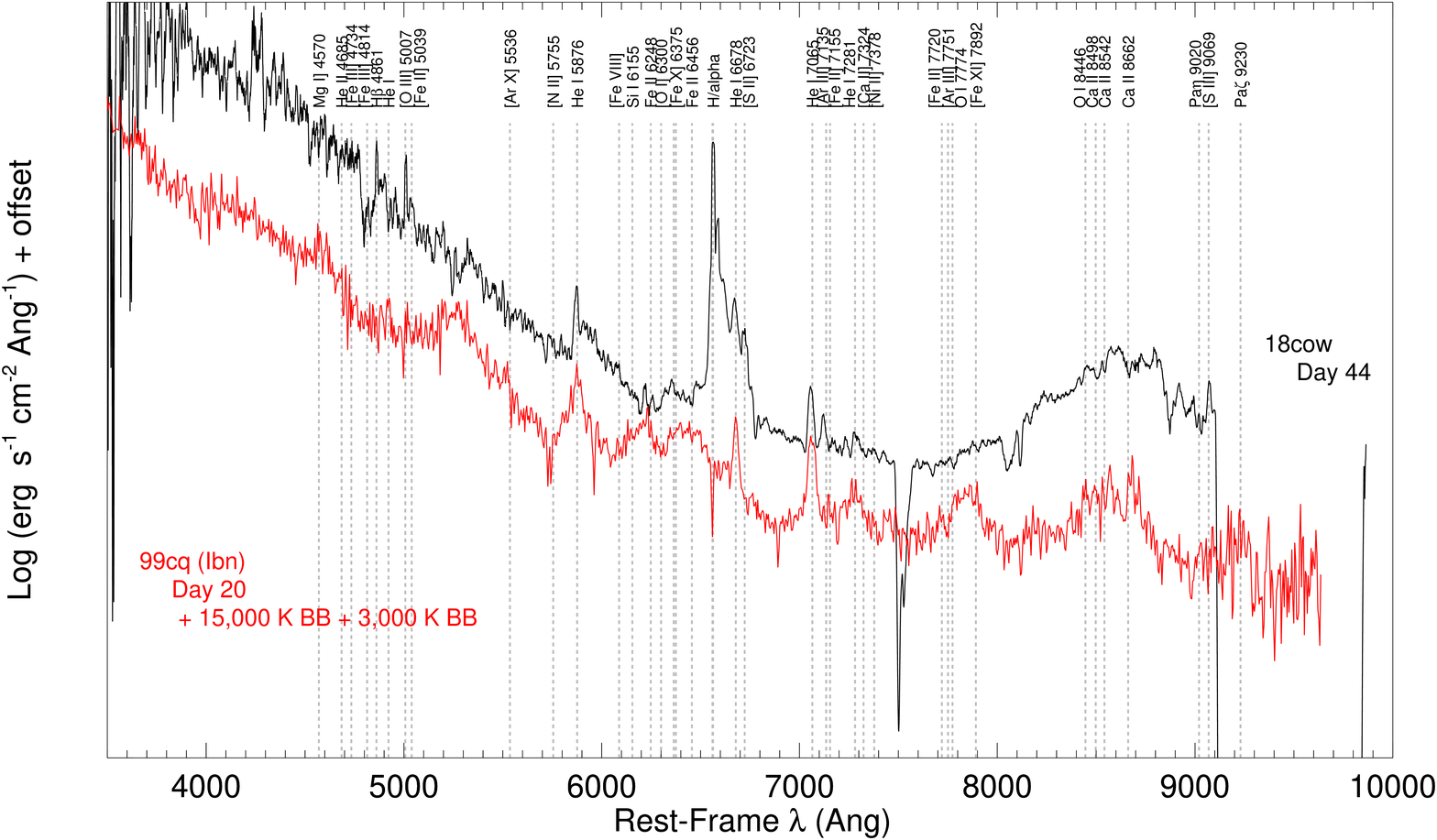}
\caption{{\bf (Top)} Comparison of AT2018cow (black) with the Type Ibn SN 1999cq (red).  {\bf (Bottom)} The spectrum of SN 1999cq is combined with 15,000 K and 3,000 K blackbody spectra that represents the optically thick photosphere and an additional dust component.  Many high-ionization spectral lines typically seen in interacting SNe are labeled, but not all are necessarily identified in each spectrum.}
\label{fig6} 
\end{figure*}

\begin{figure*}
\centering
\includegraphics[width=7.5in]{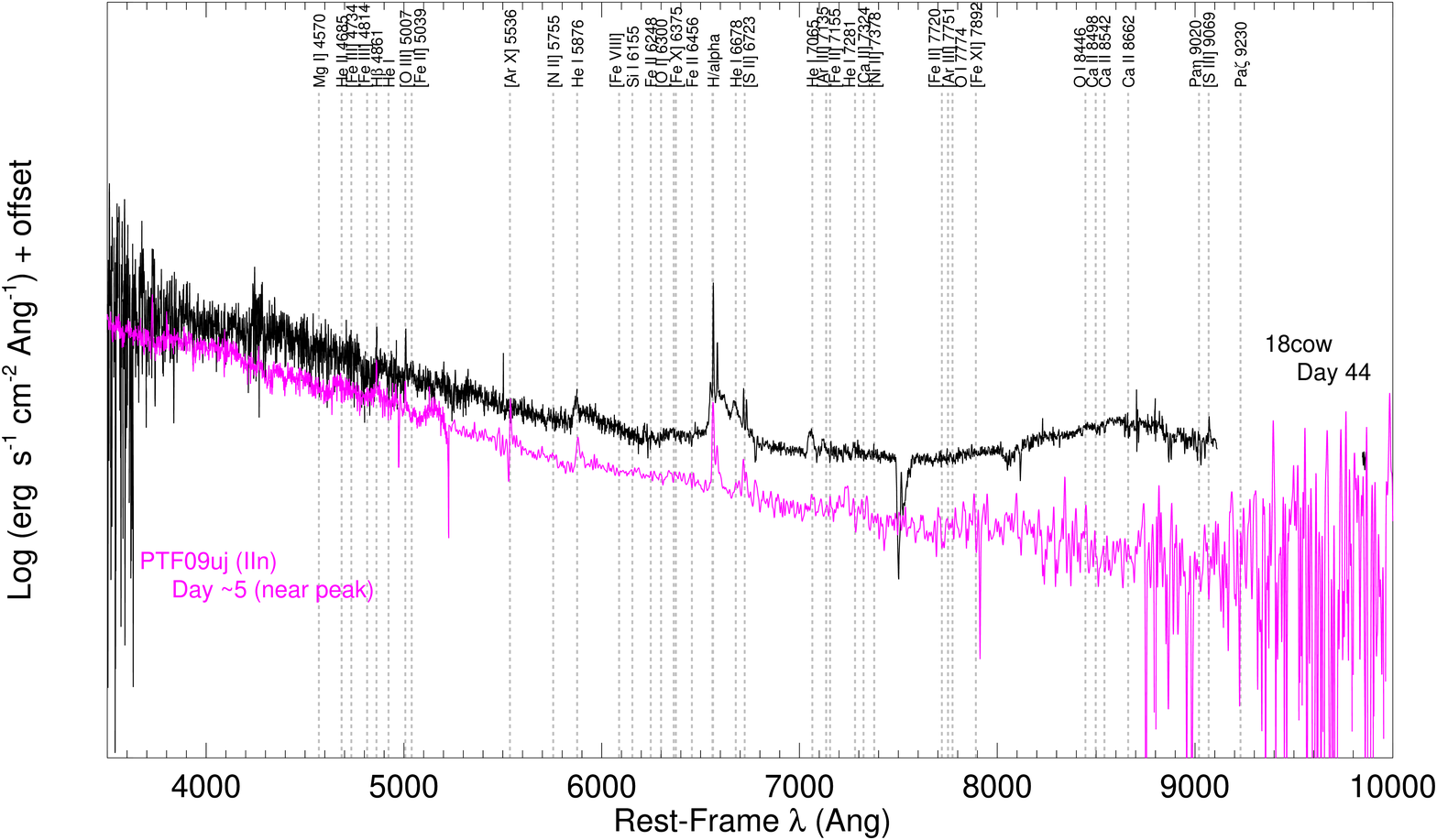}
\caption{Comparison of AT2018cow (black) with the Type IIn PTF09uj (magenta).  Many high-ionization spectral lines typically seen in interacting SNe are labeled, but not all are necessarily identified in each spectrum.}
\label{fig7} 
\end{figure*}

Figures \ref{fig3}, \ref{fig4}, \ref{fig5}, \ref{fig6}, and \ref{fig7} compare day 44 and 53 spectra of AT2018cow from \citet{perley18} to SNe 1998S (IIn), 2006jc (Ibn), 2011fw (IIn/Inb), 1999cq (Ibn), and PTF09uj (IIn?), respectively \footnote{All spectra besides AT2018cow were downloaded from either the Berkeley SN Database (http://heracles.astro.berkeley.edu/sndb/), the PTF Marshall (http://ptf.caltech.edu/cgi-bin/ptf/transient/marshal.cgi), or the Transient Name Server (https://wis-tns.weizmann.ac.il)}.  We do not have any spectra of AT208cow at epochs later than day 53.  SN 1998S is one of the most well-studied SNe IIn \citep{leonard00,shivvers15,fassia01} and used for comparison in \citet{perley18}.  We note, however, that SN 1998S may not be a prototypical member of the Type IIn subclass \citep{shivvers15,smith17,smith17b}.  The narrow lines disappeared after only a few weeks.  The SN then had a more normal ejecta-dominated spectrum with P Cygni lines.  SN 2006jc is one of the most well-studied SNe Ibn \citep{foley07,pastorello07,smith08jc}.  SN 2011hw is a transitional Type IIn/Inb that serves as a useful counterpoint to SN 2006jc \citep{smith12hw}.  We do not include SN 2011hw in our other phase space plots because the explosion date, rise time, and actual peak are not well constrained.  The observed peak luminosity of SN 2011hw is -19.5 and a decay rate of $\sim$0.05 mag/day, which is at least comparable to other members of this subclass (see Figure \ref{fig2}).  SN 1999cq is a less well-studied Type Ibn and has only one associated spectrum, but is even more extreme (in terms of rise team and peak luminosity) than AT2018cow (see Figure \ref{fig2}; \citealt{matheson00}).  PTF09uj is a peculiar Type IIn that exhibited evidence for a shock breakout from a dense circumstellar wind \citep{ofek10}.  Except when stated otherwise, all spectra are over-plotted on various scales to accentuate particular features.  Except for Figure \ref{fig7}, the spectra of AT2018cow are smoothed by 10 pixels to highlight the broader features.  Section \ref{sec:narrow} focuses on the narrower features.

Figure \ref{fig3} first compares AT2018cow (days 44 and 53) to the Type IIn SN 1998S (days 24 and 107).  As \citet{perley18} point out, at early times both AT2018cow and SN 1998S are dominated by a hot, blue continuum from the photosphere.  They also share similar spectral lines, particularly H and He.  \citet{perley18} point out that SN 1998S has much narrower lines than AT2018cow (a few hundred km/s versus $v\sim6000$ km/s), but we note that the narrow (e.g., few hundred km/s) lines in AT2018cow are also likely associated with an excited, pre-shocked CSM and not the underlying HII region.

Figure \ref{fig3} shows that after a couple months, the narrow lines in SN 1998S faded and the spectra begin to show a normal ejecta ($\sim$10,000 km/s) dominated photosphere.  (After the continuum photosphere fades in SN 1998S, the spectra again go on to show intermediate lines from CSM interaction at later times, such as in PTF11iqb \citep{smith15iqb}).  By day 44 and 53, the spectra of AT2018cow already begin to show evidence for broader, underlying components that are likely associated with the ejecta.  In most cases, these broad components appear noticeably redshifted.

Figure \ref{fig4} compares AT2018cow (days 44 and 53) to the Type Ibn SN 2006jc (day 72).  Overall, AT2018cow is bluer, but SN 2006jc shows line widths that are more comparable to AT2018cow ($1000 < v < 4000$~km/s).  The line widths in SN 2006jc are attributed to post-shocked CSM \citep{smith08jc}.  SN 2006jc also has very little helium compared to most SNe IIn, including SN 1998S.  The intermediate lines of AT2018cow may likely be associated with post-shocked CSM, too (note that the very narrow features are likely associated with the underlying galaxy).  AT2018cow also develops other lines that are typically observed in SNe, including O I and Ca II.

AT2018cow and SN 2006jc have similar spectral line identifications, although AT2018cow has significant emission just long-ward of the Ca II triplet that we cannot identify.  The AT2018cow spectral lines are also a bit weaker than SN 2006jc.  \citet{leloudas15} point out that an underlying spectrum can often be suppressed by ejecta strongly interacting with the surrounding CSM.  We therefore take a blackbody to model the hot 15,000 blackbody continuum \citep{perley18} and add this to the spectrum of SN 2006jc (Figure \ref{fig4}).  While only a qualitative argument, this model is meant to illustrate the effects of an underlying, hot photosphere.  The shape of the new SN 2006jc spectrum is comparable to AT2018cow and the H and He features are suppressed. 

Figure \ref{fig5} compares AT2018cow to similar resolution optical spectra of the transitional Type IIn/Ibn object SN~2011hw (days 7 and 47 post-{\it discovery}).  As with SN 2006jc, we add a 15,000 K blackbody.  \citet{smith12hw} adopted a Galactic extinction of E(B-V)=0.115, but note they did not account for local extinction.  We invoke additional local extinction and deredden our spectrum by a value E(B-V)=0.415. These modifications serve to make the continuum slope comparable in SN~2011hw and AT2018cow.   Compared to SN~2006jc, SN 2011hw has narrower and weaker He I lines from the CSM, and relatively stronger H$\alpha$ emission, helping to illustrate the diversity among known examples of SNe Ibn.   In terms of narrow line widths and strengths, the spectrum of AT2018cow has more in common with SN~2011hw than with SN~2006jc.  SN 2011hw, however, lacks the very strong and broad CaII IR triplet emission fromt eh underlying SN ejecta.  For SN 2011hw, the slower CSM speed and apparently higher relative H abundance might suggest that its progenitor was in a transitional evolutionary state between the RSG or LBV-like progenitors of SNe~IIn and the more Wolf-Rayet-like progenitors of SNe~Ibn.  This may apply to AT~2018cow as well, although th erelatively strong H$\alpha$ emission from a coincidence H~{\sc ii} region makes it difficult to determine if there is also some weak narrow H$\alpha$ that may arise in the CSM.

Figure \ref{fig6} goes on to make a similar comparison to the Type Ibn 1999cq (day 20), which may be a better comparison because it was almost identical to AT2018cow in terms of peak magnitude and rise time (Figure \ref{fig2}).  Despite the similarities in Figure \ref{fig2}, the spectral shapes are not a good qualitative match, but the line identifications, widths, and strengths of AT2018cow appear to be much more similar to SN 1999cq than SN 2006jc.  Like with SN 2006jc, we add a 15,000 K blackbody.  We also add a second blackbody (3,000 K) to represent hot dust \citep{perley18}.  The qualitative similarities are intriguing.  SN 1999cq even has the same emission long-ward of Ca II, which may just be very broad and/or red-shifted Ca II.

Finally, Figure \ref{fig7} compares the day 44 spectrum of AT2018cow to the near-peak spectrum of PTF09uj.  The same scaling is used and neither spectrum is smoothed.  The shape, color, line strength, and line widths show a remarkable comparison.

\subsection{X-Ray Light Curve}
\label{sec:xray}

\begin{figure}
\centering
\includegraphics[width=3.6in]{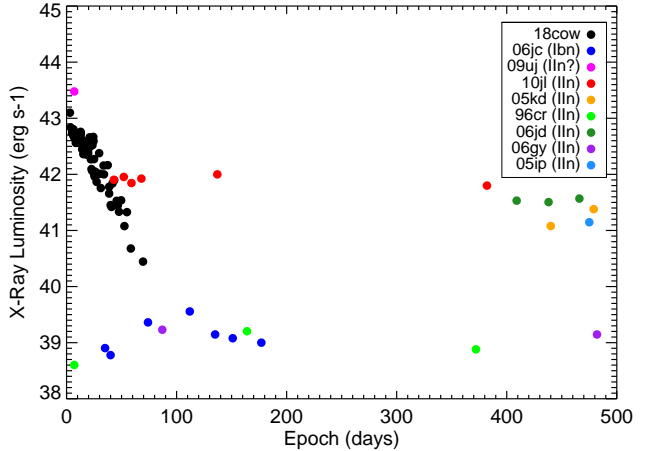}
\caption{X-Ray Light Curve of AT2018cow compared to other SNe IIn and the Type Ibn SN 2006jc.  Note that the PTF09uj photometry is near-UV from GALEX.}
\label{fig8} 
\end{figure}

\citet{margutti18} plot the X-ray light-curve of AT2018cow and identify a change in the slope at $\sim$20 days.  Overall, they conclude the multi-wavelength properties of AT2018cow to be most consistent with a central powering source, such as a magnetar.  Figure \ref{fig8} compares the X-ray light-curve of AT2018cow with other interacting SNe with X-ray observations at $<$500 days \footnote{Data were downloaded from the SN X-ray database at http://kronos.uchicago.edu/snax/}.  The only existing data that fit these criteria include the Type Ibn SN 2006jc \citep{ofek13}, and Type IIn SNe 2010jl \citep{chandra12,chandra15}, PTF09uj \citep{ofek10}, 2005kd \citep{immler07,dwarkadas16}, 1996cr \citep{dwarkadas10}, 2006jd \citep{katsuda16}, 2006gy \citep{smith07gy}, and 2005ip \citep{katsuda14}.  Only SNe 2006jc, 2010jl, and 1996cr have X-ray observations at $<$60 days (the PTF09uj photometry is near-UV from GALEX).

PTF09uj has a single early time UV data point that is slightly larger than AT2018cow and was attributed to shock breakout from a dense CSM.  In the X-rays, SN 2010jl is $\sim100\times$~fainter than the peak of AT2018cow, but the initial X-ray observations of SN 2010jl were not obtained until more than 50 days post-explosion.  By this epoch, the X-ray luminosity of SN 2010jl is similar to AT2018cow.  The X-ray luminosity of SN 2010jl is attributed to shock interaction with a dense CSM that extends out to large radii.  For this reason, the X-ray light-curve for SN 2010jl plateaus for years.  The Type Ibn SN 2006jc is significantly less luminous than AT2018cow at early-times.  As noted above, however, the shock did not hit the SN 2006jc shell until $\sim$50 days or so after peak.  At these epochs, the X-ray light-curves of AT2018cow and SN 2006jc are quite similar.  

\section{Conclusion}
\label{sec:conclusion}

This article explores the hypothesis that CSM interaction arising from strong mass loss of a H-depleted progenitor might be relevant in explaining the observed features of AT2018cow.  We note that some of the narrow lines in spectra of AT2018cow, namely the He I lines at 5876, 6680, and 7065 \AA, may trace true pre-shock CSM expanding slowly away from the progenitor star, contrary to previous suggestions.  Thee He I lines are in fact resolved, with width of around 300 km s$^{-1}$ or more, whereas other lines such as [O III] and [S II] in the same exposures are more than 2 times narrower and seem consistent with unresolved H~II region emission.  We compare AT2018cow to a number of different interacting SNe, including several SNe Ibn and IIn.  Although none of the objects provide a perfect match, many similarities exist between AT2018cow and SNe Ibn, including light-curve rise and fall times, peak magnitude, X-ray light-curves, and spectroscopic properties.  CSM interaction can be quite diverse, depending on the extent and geometry of the CSM, which is primarily governed by the timing of the pre-SN mass-loss.  Not all interacting SNe need to look exactly like SN 2006jc or SN 1998S.  Based on the optical and X-ray light curves and the spectral evolution, we infer that the interaction with H-depleted CSM in AT2018cow likely occurred earlier than in the case of SN 2006jc.

CSM interaction may also help explain two other notable features of AT2018cow.  First, the growing infrared component at $>30$~days \citep{perley18} can be understood in the context of dust formation, similar to the day $>$50 post-shock dust formation in SN 2006jc \citep{smith08jc}.   Second,  the hard and soft X-ray components observed by \citet{margutti18} are common in SNe exhibiting CSM interaction, such as SNe IIn 2005kd, 2006jd, and 2010jl \citep{katsuda16}.  The time evolution of the X-ray emission would suggest that the most intense period of CSM interaction occured early in AT2018cow, near the time of peak optical luminosity.  This, in tunrs, would imply a relatively compact, recently ejected CSM shell.

The host galaxy of AT2018cow is a dwarf spiral galaxy with a star formation rate of 0.22 \msolar~yr$^{-1}$ \citep{perley18}.  This rate is consistent with the hosts of most core-collapse SNe, including SNe Ibn \citep{hosseinzadeh19}.  \citet{perley18} point out, however, that the galaxy mass ($1.4\times10^9$~\msolar) is smaller than a majority of galaxies that produce core-collapse SNe and that the galaxy is not undergoing a notable burst of star-formation at the current time.  The initial classification of AT2018cow as a magnetar in a dwarf spiral galaxy may be reminiscent of the initial classification of the Type Ibn PS1-12sk \citep{sanders13}.  The host was initially identified as brightest cluster galaxy (BCG) CGCG 208–042, but there is little to no star-formation at the SN location.  \citet{hosseinzadeh19} suggest that the actual galaxy may be the more nearby compact dwarf galaxy 2.7 kpc away or, more likely, the progenitor of PS1-12sk was not a massive star.  Instead, they propose that the progenitor could be white dwarf in a helium envelope or a non-degenerate helium star.  This hypothesis is also consistent with the electron-capture collapse following a merger of a massive ONeMg white dwarf (WD) with another WD proposed for FBOTs and AT2018cow \citep{lyutikov18}.

The nature of SN Ibn progenitors remains uncertain.  Some SNe Ibn (e.g., SN 2006jc) likely have massive progenitors \citep{foley07,pastorello07}, but it is unclear if this applies to all examples.  Given some of the observational similarities highlighted in this paper, it is tempting to consider a connection between AT2018cow (and perhaps all FBOTs) and the Type~Ibn SN subclass.  For the relatively quick timescales observed in these objects, any CSM would need to be quite localized in extent.  Some models propose eruptive mass loss on short timescales, even in hydrogen poor stars \citep[e.g.,][]{quataert12,smitharnett14,fuller15}.  Variations in timing, mass, expansion speed, and duration of pre-SN mass loss might then result in the observational differences discussed throughout this article.  For the specific case of AT2018cow, for example, a He-rich shell much closer to the star (as compared to SN 2006jc) might result in an optical depth that is too high for strong He I line emission at early times (producing instead a hot blackbody-like optical spectrum).  The strong X-rays initially followed by a rapid drop might be consistent with this compact shell and high optical depths if the CSM is clumpy or if the CSM is globally asymmetric, allowing the X-rays to partially escape as inferred for some SNe IIn like SN 2005ip \citep{smith09ip}.  A relatively compact shell with a modest total mass would yield a faster diffusion time (e.g., rapid UV/optical rise and fall), as compared to the  more extended shells of typical SNe IIn.   In SN 2006jc, the intermediate He I lines were seen at the time of discovery (although note that SN~2006jc was discovered relatively late during its decline from peak) because they were photoionized and the CSM was relatively fast, but the SN forward shock did not collide with the densest parts of the shell until $\sim$50 days after discovery.  In contrast, by day 50, AT2018cow may have already overrun most of its CSM shell, resulting in the relatively weak narrow lines from any remaining pre-shock CSM and moderate strength intermediate-width lines from the swept-up CSM.

The observational degeneracies between TDEs, magnetars, and SNe are exacerbated by the hot blue continuum at early times.  While the diversity of CSM interaction and the relatively small number of SNe Ibn make it difficult to identify an example that matches the peculiar event AT~2018cow, there is partial similarity in their light curves and spectra.  The resolved intermediate-width and narrow components of He I lines, in particular, make a strong case that CSM interaction plays an important role in AT~2018cow.    At later times, if CSM interaction continues or if the swept up CSM continues to emit, the light-curve may plateau or intermediate-width lines may become more dominant in the spectrum. AT2018cow is currently behind the Sun.  When it emerges, late-time, multi-wavelength observations may be able to distinguish between various scenarios once the photospheric continuum has faded, perhaps allowing CSM interaction signatures to be identified more clearly. 
\newline
\newline


\noindent{\bf Acknowledgements}

The authors wish to thank Maxim Lyutikov and Sarah Doverspike for useful conversations.

\bibliographystyle{apj2}
\bibliography{references2}

\begin{thebibliography}{}
\expandafter\ifx\csname natexlab\endcsname\relax\def\natexlab#1{#1}\fi

\bibitem[{{Arcavi} {et~al.}(2016){Arcavi}, {Wolf}, {Howell}, {Bildsten},
  {Leloudas}, {Hardin}, {Prajs}, {Perley}, {Svirski}, {Gal-Yam}, {Katz},
  {McCully}, {Cenko}, {Lidman}, {Sullivan}, {Valenti}, {Astier}, {Balland},
  {Carlberg}, {Conley}, {Fouchez}, {Guy}, {Pain}, {Palanque-Delabrouille},
  {Perrett}, {Pritchet}, {Regnault}, {Rich}, \& {Ruhlmann-Kleider}}]{arcavi16}
{Arcavi}, I., {Wolf}, W.~M., {Howell}, D.~A., {et~al.} 2016, ApJ, 819, 35

\bibitem[{Balberg \& Loeb(2011)}]{balberg11}
Balberg, S., \& Loeb, A. 2011, MNRAS, 414, 1715

\bibitem[{{Bilinski} {et~al.}(2018){Bilinski}, {Smith}, {Williams}, {Smith},
  {Zheng}, {Graham}, {Mauerhan}, {Andrews}, {Filippenko}, {Akerlof},
  {Chatzopoulos}, {Hoffman}, {Huk}, {Leonard}, {Marion}, {Milne}, {Quimby},
  {Silverman}, {Vink{\'o}}, {Wheeler}, \& {Yuan}}]{bilinski18}
{Bilinski}, C., {Smith}, N., {Williams}, G.~G., {et~al.} 2018, MNRAS, 475, 1104

\bibitem[{{Chandra} {et~al.}(2012){Chandra}, {Chevalier}, {Chugai}, {Fransson},
  {Irwin}, {Soderberg}, {Chakraborti}, \& {Immler}}]{chandra12}
{Chandra}, P., {Chevalier}, R.~A., {Chugai}, N., {et~al.} 2012, ApJ, 755, 110

\bibitem[{{Chandra} {et~al.}(2015){Chandra}, {Chevalier}, {Chugai}, {Fransson},
  \& {Soderberg}}]{chandra15}
{Chandra}, P., {Chevalier}, R.~A., {Chugai}, N., {Fransson}, C., \&
  {Soderberg}, A.~M. 2015, ApJ, 810, 32

\bibitem[{Chevalier \& Irwin(2011)}]{chevalier11}
Chevalier, R.~A., \& Irwin, C.~M. 2011, ApJL, 729, L6

\bibitem[{{Drout} {et~al.}(2014){Drout}, {Chornock}, {Soderberg}, {Sanders},
  {McKinnon}, {Rest}, {Foley}, {Milisavljevic}, {Margutti}, {Berger},
  {Calkins}, {Fong}, {Gezari}, {Huber}, {Kankare}, {Kirshner}, {Leibler},
  {Lunnan}, {Mattila}, {Marion}, {Narayan}, {Riess}, {Roth}, {Scolnic},
  {Smartt}, {Tonry}, {Burgett}, {Chambers}, {Hodapp}, {Jedicke}, {Kaiser},
  {Magnier}, {Metcalfe}, {Morgan}, {Price}, \& {Waters}}]{drout14}
{Drout}, M.~R., {Chornock}, R., {Soderberg}, A.~M., {et~al.} 2014, ApJ, 794, 23

\bibitem[{{Dwarkadas} {et~al.}(2010){Dwarkadas}, {Dewey}, \&
  {Bauer}}]{dwarkadas10}
{Dwarkadas}, V.~V., {Dewey}, D., \& {Bauer}, F. 2010, MNRAS, 407, 812

\bibitem[{{Dwarkadas} {et~al.}(2016){Dwarkadas}, {Romero-Ca{\~n}izales},
  {Reddy}, \& {Bauer}}]{dwarkadas16}
{Dwarkadas}, V.~V., {Romero-Ca{\~n}izales}, C., {Reddy}, R., \& {Bauer}, F.~E.
  2016, MNRAS, 462, 1101

\bibitem[{{Esteban} {et~al.}(2004){Esteban}, {Peimbert}, {Garc{\'{\i}}a-Rojas},
  {Ruiz}, {Peimbert}, \& {Rodr{\'{\i}}guez}}]{esteban04}
{Esteban}, C., {Peimbert}, M., {Garc{\'{\i}}a-Rojas}, J., {et~al.} 2004, MNRAS,
  355, 229

\bibitem[{Fassia {et~al.}(2001)Fassia, Meikle, Chugai, Geballe, Lundqvist,
  Walton, Pollacco, Veilleux, Wright, Pettini, Kerr, Puchnarewicz, Puxley,
  Irwin, Packham, Smartt, \& Harmer}]{fassia01}
Fassia, A., Meikle, W. P.~S., Chugai, N., {et~al.} 2001, Monthly Notices RAS,
  325, 907

\bibitem[{Foley {et~al.}(2007)Foley, Smith, Ganeshalingam, Li, Chornock, \&
  Filippenko}]{foley07}
Foley, R.~J., Smith, N., Ganeshalingam, M., {et~al.} 2007, ApJL, 657, L105

\bibitem[{{Fuller} {et~al.}(2015){Fuller}, {Cantiello}, {Lecoanet}, \&
  {Quataert}}]{fuller15}
{Fuller}, J., {Cantiello}, M., {Lecoanet}, D., \& {Quataert}, E. 2015, ApJ,
  810, 101

\bibitem[{{Ginzburg} \& {Balberg}(2014)}]{ginzburg14}
{Ginzburg}, S., \& {Balberg}, S. 2014, ApJ, 780, 18

\bibitem[{{Hosseinzadeh} {et~al.}(2019){Hosseinzadeh}, {McCully}, {Zabludoff},
  {Arcavi}, {French}, {Howell}, {Berger}, \& {Hiramatsu}}]{hosseinzadeh19}
{Hosseinzadeh}, G., {McCully}, C., {Zabludoff}, A.~I., {et~al.} 2019, arXiv
  e-prints, arXiv:1901.03332

\bibitem[{{Hosseinzadeh} {et~al.}(2017){Hosseinzadeh}, {Arcavi}, {Valenti},
  {McCully}, {Howell}, {Johansson}, {Sollerman}, {Pastorello}, {Benetti},
  {Cao}, {Cenko}, {Clubb}, {Corsi}, {Duggan}, {Elias-Rosa}, {Filippenko},
  {Fox}, {Fremling}, {Horesh}, {Karamehmetoglu}, {Kasliwal}, {Marion}, {Ofek},
  {Sand}, {Taddia}, {Zheng}, {Fraser}, {Gal-Yam}, {Inserra}, {Laher}, {Masci},
  {Rebbapragada}, {Smartt}, {Smith}, {Sullivan}, {Surace}, \&
  {Wo{\'z}niak}}]{hosseinzadeh17}
{Hosseinzadeh}, G., {Arcavi}, I., {Valenti}, S., {et~al.} 2017, ApJ, 836, 158

\bibitem[{Immler \& Pooley(2007)}]{immler07}
Immler, S., \& Pooley, D. 2007, ATEL, 1004, 1

\bibitem[{Kasen \& Bildsten(2010)}]{kasen10}
Kasen, D., \& Bildsten, L. 2010, ApJ, 717, 245

\bibitem[{Katsuda {et~al.}(2014)Katsuda, Maeda, Nozawa, Pooley, \&
  Immler}]{katsuda14}
Katsuda, S., Maeda, K., Nozawa, T., Pooley, D., \& Immler, S. 2014, The
  Astrophysical Journal, 780, 184

\bibitem[{{Katsuda} {et~al.}(2016){Katsuda}, {Maeda}, {Bamba}, {Terada},
  {Fukazawa}, {Kawabata}, {Ohno}, {Sugawara}, {Tsuboi}, \&
  {Immler}}]{katsuda16}
{Katsuda}, S., {Maeda}, K., {Bamba}, A., {et~al.} 2016, ApJ, 832, 194

\bibitem[{Kiewe {et~al.}(2012)Kiewe, Gal-Yam, Arcavi, Leonard, Enriquez, Cenko,
  Fox, Moon, Sand, Soderberg, \& CCCP}]{kiewe12}
Kiewe, M., Gal-Yam, A., Arcavi, I., {et~al.} 2012, ApJ, 744, 10

\bibitem[{{Kuin} {et~al.}(2018){Kuin}, {Wu}, {Oates}, {Lien}, {Emery},
  {Kennea}, {de Pasquale}, {Han}, {Brown}, {Tohuvavohu}, {Breeveld}, {Burrows},
  {Cenko}, {Campana}, {Levan}, {Markwardt}, {Osborne}, {Page}, {Page},
  {Sbarufatti}, {Siegel}, \& {Troja}}]{kuin18}
{Kuin}, N.~P.~M., {Wu}, K., {Oates}, S., {et~al.} 2018, arXiv e-prints,
  arXiv:1808.08492

\bibitem[{{Leloudas} {et~al.}(2015){Leloudas}, {Hsiao}, {Johansson}, {Maeda},
  {Moriya}, {Nordin}, {Petrushevska}, {Silverman}, {Sollerman}, {Stritzinger},
  {Taddia}, \& {Xu}}]{leloudas15}
{Leloudas}, G., {Hsiao}, E.~Y., {Johansson}, J., {et~al.} 2015, A\&A, 574, A61

\bibitem[{Leonard {et~al.}(2000)Leonard, Filippenko, Barth, \&
  Matheson}]{leonard00}
Leonard, D.~C., Filippenko, A.~V., Barth, A.~J., \& Matheson, T. 2000, The
  Astrophysical Journal, 536, 239

\bibitem[{{Lyutikov} \& {Toonen}(2018)}]{lyutikov18}
{Lyutikov}, M., \& {Toonen}, S. 2018, arXiv e-prints, arXiv:1812.07569

\bibitem[{{Margutti} {et~al.}(2018){Margutti}, {Metzger}, {Chornock}, {Vurm},
  {Roth}, {Grefenstette}, {Savchenko}, {Cartier}, {Steiner}, {Terreran},
  {Migliori}, {Milisavljevic}, {Alexander}, {Bietenholz}, {Blanchard}, {Bozzo},
  {Brethauer}, {Chilingarian}, {Coppejans}, {Ducci}, {Ferrigno}, {Fong},
  {G{\"O}tz}, {Guidorzi}, {Hajela}, {Hurley}, {Kuulkers}, {Laurent},
  {Mereghetti}, {Nicholl}, {Patnaude}, {Ubertini}, {Banovetz}, {Bartel},
  {Berger}, {Coughlin}, {Eftekhari}, {Frederiks}, {Kozlova}, {Laskar},
  {Svinkin}, {Drout}, {Macfadyen}, \& {Paterson}}]{margutti18}
{Margutti}, R., {Metzger}, B.~D., {Chornock}, R., {et~al.} 2018, arXiv
  e-prints, arXiv:1810.10720

\bibitem[{Matheson {et~al.}(2000)Matheson, Filippenko, Ho, Barth, \&
  Leonard}]{matheson00}
Matheson, T., Filippenko, A.~V., Ho, L.~C., Barth, A.~J., \& Leonard, D.~C.
  2000, AJ, 120, 1499

\bibitem[{{Morozova} {et~al.}(2018){Morozova}, {Piro}, \&
  {Valenti}}]{morozova18}
{Morozova}, V., {Piro}, A.~L., \& {Valenti}, S. 2018, ApJ, 858, 15

\bibitem[{{Ofek} {et~al.}(2010){Ofek}, {Rabinak}, {Neill}, {Arcavi}, {Cenko},
  {Waxman}, {Kulkarni}, {Gal-Yam}, {Nugent}, {Bildsten}, {Bloom}, {Filippenko},
  {Forster}, {Howell}, {Jacobsen}, {Kasliwal}, {Law}, {Martin}, {Poznanski},
  {Quimby}, {Shen}, {Sullivan}, {Dekany}, {Rahmer}, {Hale}, {Smith},
  {Zolkower}, {Velur}, {Walters}, {Henning}, {Bui}, \& {McKenna}}]{ofek10}
{Ofek}, E.~O., {Rabinak}, I., {Neill}, J.~D., {et~al.} 2010, ApJ, 724, 1396

\bibitem[{{Ofek} {et~al.}(2013){Ofek}, {Fox}, {Cenko}, {Sullivan}, {Gnat},
  {Frail}, {Horesh}, {Corsi}, {Quimby}, {Gehrels}, {Kulkarni}, {Gal-Yam},
  {Nugent}, {Yaron}, {Filippenko}, {Kasliwal}, {Bildsten}, {Bloom},
  {Poznanski}, {Arcavi}, {Laher}, {Levitan}, {Sesar}, \& {Surace}}]{ofek13}
{Ofek}, E.~O., {Fox}, D., {Cenko}, S.~B., {et~al.} 2013, ApJ, 763, 42

\bibitem[{{Pastorello} {et~al.}(2007){Pastorello}, {Smartt}, {Mattila},
  {Eldridge}, {Young}, {Itagaki}, {Yamaoka}, {Navasardyan}, {Valenti}, {Patat},
  {Agnoletto}, {Augusteijn}, {Benetti}, {Cappellaro}, {Boles}, {Bonnet-Bidaud},
  {Botticella}, {Bufano}, {Cao}, {Deng}, {Dennefeld}, {Elias-Rosa},
  {Harutyunyan}, {Keenan}, {Iijima}, {Lorenzi}, {Mazzali}, {Meng}, {Nakano},
  {Nielsen}, {Smoker}, {Stanishev}, {Turatto}, {Xu}, \&
  {Zampieri}}]{pastorello07}
{Pastorello}, A., {Smartt}, S.~J., {Mattila}, S., {et~al.} 2007, Nature, 447,
  829

\bibitem[{{Peimbert} {et~al.}(2000){Peimbert}, {Peimbert}, \&
  {Ruiz}}]{peimbert00}
{Peimbert}, M., {Peimbert}, A., \& {Ruiz}, M.~T. 2000, ApJ, 541, 688

\bibitem[{{Perley} {et~al.}(2018){Perley}, {Mazzali}, {Yan}, {Cenko}, {Gezari},
  {Taggart}, {Blagorodnova}, {Fremling}, {Mockler}, {Singh}, {Tominaga},
  {Tanaka}, {Watson}, {Ahumada}, {Anupama}, {Ashall}, {Becerra}, {Bersier},
  {Bhalerao}, {Bloom}, {Butler}, {Copperwheat}, {Coughlin}, {De}, {Drake},
  {Duev}, {Frederick}, {Jes{\'u}s Gonz{\'a}lez}, {Goobar}, {Heida}, {Ho},
  {Horst}, {Hung}, {Itoh}, {Jencson}, {Kasliwal}, {Kawai}, {Kulkarni}, {Kumar},
  {Kumar}, {Kutyrev}, {Khanam}, {Lee}, {Maeda}, {Mahabal}, {Murata}, {Neill},
  {Ngeow}, {Penprase}, {Pian}, {Quimby}, {Ramirez-Ruiz}, {Richer},
  {Rom{\'a}n-Z{\'u}{\~n}iga}, {Srivastava}, {Socia}, {Sollerman}, {Tachibana},
  {Taddia}, {Tinyanont}, {Troja}, {Ward}, \& {Wee}}]{perley18}
{Perley}, D.~A., {Mazzali}, P.~A., {Yan}, L., {et~al.} 2018, arXiv e-prints,
  arXiv:1808.00969

\bibitem[{{Prentice} {et~al.}(2018){Prentice}, {Maguire}, {Smartt}, {Magee},
  {Schady}, {Sim}, {Chen}, {Clark}, {Colin}, {Fulton}, {McBrien}, {ONeill},
  {Smith}, {Ashall}, {Chambers}, {Denneau}, {Flewelling}, {Heinze}, {Holoien},
  {Huber}, {Kochanek}, {Mazzali}, {Prieto}, {Rest}, {Shappee}, {Stalder},
  {Stanek}, {Stritzinger}, {Thompson}, \& {Tonry}}]{prentice18}
{Prentice}, S.~J., {Maguire}, K., {Smartt}, S.~J., {et~al.} 2018, ApJL, 865, L3

\bibitem[{{Pursiainen} {et~al.}(2018){Pursiainen}, {Childress}, {Smith},
  {Prajs}, {Sullivan}, {Davis}, {Foley}, {Asorey}, {Calcino}, {Carollo},
  {Curtin}, {D'Andrea}, {Glazebrook}, {Gutierrez}, {Hinton}, {Hoormann},
  {Inserra}, {Kessler}, {King}, {Kuehn}, {Lewis}, {Lidman}, {Macaulay},
  {M{\"o}ller}, {Nichol}, {Sako}, {Sommer}, {Swann}, {Tucker}, {Uddin},
  {Wiseman}, {Zhang}, {Abbott}, {Abdalla}, {Allam}, {Annis}, {Avila}, {Brooks},
  {Buckley-Geer}, {Burke}, {Carnero Rosell}, {Carrasco Kind}, {Carretero},
  {Castander}, {Cunha}, {Davis}, {De Vicente}, {Diehl}, {Doel}, {Eifler},
  {Flaugher}, {Fosalba}, {Frieman}, {Garc{\'{\i}}a-Bellido}, {Gruen},
  {Gruendl}, {Gutierrez}, {Hartley}, {Hollowood}, {Honscheid}, {James},
  {Jeltema}, {Kuropatkin}, {Li}, {Lima}, {Maia}, {Martini}, {Menanteau},
  {Ogando}, {Plazas}, {Roodman}, {Sanchez}, {Scarpine}, {Schindler}, {Smith},
  {Soares-Santos}, {Sobreira}, {Suchyta}, {Swanson}, {Tarle}, {Tucker}, \&
  {Walker}}]{pursiainen18}
{Pursiainen}, M., {Childress}, M., {Smith}, M., {et~al.} 2018, MNRAS, 481, 894

\bibitem[{{Quataert} \& {Shiode}(2012)}]{quataert12}
{Quataert}, E., \& {Shiode}, J. 2012, MNRAS, 423, L92

\bibitem[{{Rest} {et~al.}(2018){Rest}, {Garnavich}, {Khatami}, {Kasen},
  {Tucker}, {Shaya}, {Olling}, {Mushotzky}, {Zenteno}, {Margheim},
  {Strampelli}, {James}, {Smith}, {F{\"o}rster}, \& {Villar}}]{rest18}
{Rest}, A., {Garnavich}, P.~M., {Khatami}, D., {et~al.} 2018, Nature Astronomy,
  2, 307

\bibitem[{{Rivera Sandoval} {et~al.}(2018){Rivera Sandoval}, {Maccarone},
  {Corsi}, {Brown}, {Pooley}, \& {Wheeler}}]{sandoval18}
{Rivera Sandoval}, L.~E., {Maccarone}, T.~J., {Corsi}, A., {et~al.} 2018,
  MNRAS, 480, L146

\bibitem[{{Sanders} {et~al.}(2013){Sanders}, {Soderberg}, {Foley}, {Chornock},
  {Milisavljevic}, {Margutti}, {Drout}, {Moe}, {Berger}, {Brown}, {Lunnan},
  {Smartt}, {Fraser}, {Kotak}, {Magill}, {Smith}, {Wright}, {Huang}, {Urata},
  {Mulchaey}, {Rest}, {Sand}, {Chomiuk}, {Friedman}, {Kirshner}, {Marion},
  {Tonry}, {Burgett}, {Chambers}, {Hodapp}, {Kudritzki}, \&
  {Price}}]{sanders13}
{Sanders}, N.~E., {Soderberg}, A.~M., {Foley}, R.~J., {et~al.} 2013, ApJ, 769,
  39

\bibitem[{Schlegel(1990)}]{schlegel90}
Schlegel, E.~M. 1990, MNRAS, 244, 269

\bibitem[{{Shivvers} {et~al.}(2015){Shivvers}, {Groh}, {Mauerhan}, {Fox},
  {Leonard}, \& {Filippenko}}]{shivvers15}
{Shivvers}, I., {Groh}, J.~H., {Mauerhan}, J.~C., {et~al.} 2015, ApJ, 806, 213

\bibitem[{{Shivvers} {et~al.}(2017){Shivvers}, {Modjaz}, {Zheng}, {Liu},
  {Filippenko}, {Silverman}, {Matheson}, {Pastorello}, {Graur}, {Foley},
  {Chornock}, {Smith}, {Leaman}, \& {Benetti}}]{shivvers17}
{Shivvers}, I., {Modjaz}, M., {Zheng}, W., {et~al.} 2017, PASP, 129, 054201

\bibitem[{{Smith}(2014)}]{smith14}
{Smith}, N. 2014, ARA\&A, 52, 487

\bibitem[{{Smith}(2017)}]{smith17b}
---. 2017, {Interacting Supernovae: Types IIn and Ibn}, ed. A.~W. {Alsabti} \&
  P.~{Murdin}, 403

\bibitem[{{Smith} \& {Arnett}(2014)}]{smitharnett14}
{Smith}, N., \& {Arnett}, W.~D. 2014, ApJ, 785, 82

\bibitem[{Smith {et~al.}(2008)Smith, Foley, \& Filippenko}]{smith08jc}
Smith, N., Foley, R.~J., \& Filippenko, A.~V. 2008, ApJ, 680, 568

\bibitem[{{Smith} {et~al.}(2011){Smith}, {Li}, {Filippenko}, \&
  {Chornock}}]{smith11}
{Smith}, N., {Li}, W., {Filippenko}, A.~V., \& {Chornock}, R. 2011, MNRAS, 412,
  1522

\bibitem[{{Smith} {et~al.}(2012){Smith}, {Mauerhan}, {Silverman},
  {Ganeshalingam}, {Filippenko}, {Cenko}, {Clubb}, \&
  {Kandrashoff}}]{smith12hw}
{Smith}, N., {Mauerhan}, J.~C., {Silverman}, J.~M., {et~al.} 2012, MNRAS, 426,
  1905

\bibitem[{{Smith} \& {McCray}(2007)}]{smithmccray07}
{Smith}, N., \& {McCray}, R. 2007, ApJL, 671, L17

\bibitem[{{Smith} \& {Morse}(2004)}]{smith04}
{Smith}, N., \& {Morse}, J.~A. 2004, ApJ, 605, 854

\bibitem[{Smith {et~al.}(2007)Smith, Li, Foley, Wheeler, Pooley, Chornock,
  Filippenko, Silverman, Quimby, Bloom, \& Hansen}]{smith07gy}
Smith, N., Li, W., Foley, R.~J., {et~al.} 2007, ApJ, 666, 1116

\bibitem[{Smith {et~al.}(2009)Smith, Silverman, Chornock, Filippenko, Wang, Li,
  Ganeshalingam, Foley, Rex, \& Steele}]{smith09ip}
Smith, N., Silverman, J.~M., Chornock, R., {et~al.} 2009, ApJ, 695, 1334

\bibitem[{{Smith} {et~al.}(2015){Smith}, {Mauerhan}, {Cenko}, {Kasliwal},
  {Silverman}, {Filippenko}, {Gal-Yam}, {Clubb}, {Graham}, {Leonard}, {Horst},
  {Williams}, {Andrews}, {Kulkarni}, {Nugent}, {Sullivan}, {Maguire}, {Xu}, \&
  {Ben-Ami}}]{smith15iqb}
{Smith}, N., {Mauerhan}, J.~C., {Cenko}, S.~B., {et~al.} 2015, MNRAS, 449, 1876

\bibitem[{{Smith} {et~al.}(2017){Smith}, {Kilpatrick}, {Mauerhan}, {Andrews},
  {Margutti}, {Fong}, {Graham}, {Zheng}, {Kelly}, {Filippenko}, \&
  {Fox}}]{smith17}
{Smith}, N., {Kilpatrick}, C.~D., {Mauerhan}, J.~C., {et~al.} 2017, MNRAS, 466,
  3021

\bibitem[{{Tanaka} {et~al.}(2016){Tanaka}, {Tominaga}, {Morokuma}, {Yasuda},
  {Furusawa}, {Baklanov}, {Blinnikov}, {Moriya}, {Doi}, {Jiang}, {Kato},
  {Kikuchi}, {Kuncarayakti}, {Nagao}, {Nomoto}, \& {Taniguchi}}]{tanaka16}
{Tanaka}, M., {Tominaga}, N., {Morokuma}, T., {et~al.} 2016, ApJ, 819, 5

\end{thebibliography}

\end{document}